\documentclass[11pt]{article}

\usepackage{latexsym,color,enumerate,graphicx,amsmath,amssymb,xspace,verbatim,cite}

\usepackage{microtype}

\usepackage{hyperref}

\newcommand{\BeginMyItemize}{\begin{itemize}\setlength{\itemsep}{-\parskip}}
\newcommand{\EndMyItemize}{\end{itemize}}

\newcommand{\BeginMyEnumerate}{\begin{enumerate}\setlength{\itemsep}{-\parskip}}
\newcommand{\EndMyEnumerate}{\end{enumerate}}

\newcommand{\etal}{\emph{et al.}}

\def\reals{\mathbb{R}}
\let\eps\varepsilon

\def\A{\mathcal{A}}
\def\G{\mathcal{G}}
\def\dd{{\sf dist}}
\def\Vor{\mathrm{Vor}}

\newtheorem{theorem}{Theorem}[section]
\newtheorem{lemma}[theorem]{Lemma}
\newtheorem{corollary}[theorem]{Corollary}
\newtheorem{proposition}[theorem]{Proposition}

\begin{document}

\title{Subquadratic Algorithms for Some \textsc{3Sum}-Hard Geometric Problems in the Algebraic Decision Tree Model\thanks{%
  Work by B.A. partially supported by NSF grants CCF-15-40656 and CCF-20-08551,
  and by grant~2014/170 from the US-Israel Binational Science Foundation.
  Work by M.d.B. partially supported by the Dutch Research Council (NWO) through Gravitation Grant NETWORKS (project no.~024.002.003.
  Work by J.C. partially supported by the F.R.S.-FNRS (Fonds National de la
  Recherche Scientifique) under CDR Grant J.0146.18.
  Work by E.E. partially supported by NSF CAREER under grant CCF:AF-1553354
  and by~grant~824/17 from the Israel Science Foundation.
  Work by J.I. partially supported by Fonds de la Recherche Scientifique FNRS under grant no.~MISU F 6001 1.
  Work by M.S. partially supported by ISF grant~260/18, by grant~1367/2016
  from the German-Israeli Science Foundation (GIF), and by
  Blavatnik Research Fund in Computer Science at Tel Aviv University.}}

\author{Boris Aronov\thanks{%
Tandon School of Engineering, 
New York University, Brooklyn NY, USA; {\sf boris.aronov@nyu.edu}}
% {http://orcid.org/0000-0003-3110-4702}{}%
\and
Mark de Berg\thanks{%
Eindhoven University of Technology, Eindhoven, Netherlands; {\sf m.t.d.berg@tue.nl}}
% {http://orcid.org/0000-0001-5770-3784}{}%
\and
Jean Cardinal\thanks{%
Universit\'e libre de Bruxelles (ULB), Brussels, Belgium; {\sf jcardin@ulb.ac.be}}
% {https://orcid.org/0000-0002-2312-0967}{}%
\and
Esther Ezra\thanks{%
School of Computer Science, Bar Ilan University, Ramat Gan, Israel; {\sf ezraest@cs.biu.ac.il}}
% {https://orcid.org/0000-0001-8133-1335}{}%
\and
John Iacono\thanks{%
Universit\'e libre de Bruxelles (ULB), Brussels, Belgium, and Tandon School of Engineering, New York University, Brooklyn NY, USA; {\sf john@johniacono.com}}
% {https://orcid.org/0000-0001-8885-8172}{}%
\and
Micha Sharir\thanks{%
School of Computer Science, Tel Aviv University, Tel~Aviv, Israel; {\sf michas@tau.ac.il}}
% {http://orcid.org/0000-0002-2541-3763}{}
}
  
\maketitle
  
\begin{abstract}
We present subquadratic algorithms in the algebraic decision-tree model
for several \textsc{3Sum}-hard geometric problems, all of which can be reduced to
the following question: Given two sets $A$, $B$, each consisting of $n$ pairwise
disjoint segments in the plane, and a set $C$ of $n$ triangles in the
plane, we want to count, for each triangle $\Delta\in C$, the number
of intersection points between the segments of $A$ and those
of $B$ that lie in $\Delta$. The problems considered in this paper
have been studied by Chan~(2020),
who gave algorithms that solve them, in the standard
real-RAM model, in $O((n^2/\log^2n)\log^{O(1)}\log n)$ time. We present
solutions in the algebraic decision-tree model whose cost is
$O(n^{60/31+\eps})$, for any $\eps>0$.

Our approach is based on a primal-dual range searching mechanism,
which exploits the multi-level polynomial partitioning machinery recently
developed by Agarwal, Aronov, Ezra, and Zahl~(2020).

A key step in the procedure is a variant of point location in arrangements,
say of lines in the plane, which is based solely on the \emph{order type}
of the lines, a ``handicap'' that turns out to be beneficial for speeding up our algorithm.
\end{abstract}

%-------------------------------------------------------------------------
\section{Introduction}
%-------------------------------------------------------------------------

Let $A$ and $B$ be two sets, each consisting of $n$ pairwise disjoint line segments
in the plane, and let $C$~be a set of $n$ triangles in the plane.
We study the problem of counting, for each triangle $\Delta\in C$, the number
of intersection points between the segments of $A$ and those of $B$
that lie inside $\Delta$. We refer to this problem as
\emph{within-triangle intersection-counting}.
This is one of four \textsc{3Sum}-hard problems (among many others) studied by
Chan~\cite{Chan}, all of which can be reduced to the problem just mentioned.\footnote{Chan~\cite{Chan} refers to this problem as ``triangle intersection-counting.''}
The other three problems are:\footnote{%
  The fact that these problems are \textsc{3Sum}-hard, and the connections between them, are stated in~\cite{Chan}.}
\begin{enumerate}[(i)]
\item 
\emph{Intersection of three polygons.}
Given three simple $n$-gons $A$, $B$, $C$ in the plane, determine whether
$A\cap B\cap C$ is nonempty.
\item 
\emph{Coverage by three polygons.}
Given three simple $n$-gons $A$, $B$, $C$ in the plane, determine whether
$A\cup B\cup C$ covers a given triangle $\Delta_0$.
\item 
\emph{Segment concurrency.}
Given sets $A$, $B$, $C$, each consisting of $n$ pairwise disjoint
segments in the plane,\footnote{%
  The segments of one set, say $C$, need not be pairwise disjoint.
  Although not explicitly stated, the technique in \cite{Chan} for the
  uniform model can also handle this situation.}
determine whether $A\times B\times C$ contains
a concurrent triple.
\end{enumerate}
Chan~\cite{Chan} presents slightly subquadratic algorithms for all four
problems, whose running time in the standard real-RAM model (also referred
to as the \emph{uniform} model) is $O((n^2/\log^2n)\log^{O(1)}\log n)$.
He has observed that, as already mentioned, all these problems can be reduced 
in near-linear time to the within-triangle intersection-counting problem, so it suffices to
present an efficient subquadratic solution for that problem.

We study the within-triangle intersection-counting problem in the algebraic decision-tree model.
In this model only
sign tests of polynomial inequalities of constant degree that access explicitly 
(the endpoint coordinates of) the input segments (or vertices of the input triangles) count towards the running time.
All other operations cost nothing in the model, but are assumed not to access the input
segments explicitly. Although originally introduced for establishing lower bounds~\cite{BenOr},
the algebraic decision-tree model has become a standard model for upper 
bounds too, used in the study of many problems,
including the \textsc{3Sum}-problem itself~\cite{CIO-16,ES,GS,GP,KLM}
and various \textsc{3Sum}-hard geometric problems~\cite{AES,BCILOS,ES}.
One can interpret the decision-tree model as an attempt to isolate and minimize the cost of 
the part of the algorithm that explicitly accesses the real representation of the input objects, 
and ignore the cost of the other purely discrete steps. This has the potential of providing us with an insight about the
problem complexity, which might eventually lead to an improved solution also in the uniform real-RAM model.

We show that the within-triangle intersection-counting problem and, hence, also
problems (i)--(iii), can be solved in this model with $O(n^{60/31+\eps})$
sign tests, for any $\eps>0$. Chan~\cite{Chan} also remarks (without providing
details) that his algorithm can be implemented in $O(n^{2-\delta})$ time in the
algebraic decision-tree model, for some $\delta>0$ that he left unspecified.
(With some care, as was communicated to us, one can obtain $\delta\approx 0.01$.)
Our algorithm is rather different from Chan's, 
and gives the concrete value $\delta=2/31$, as mentioned above.
Our techniques appear to be of independent interest and to have the potential
to apply to other problems, as we demonstrate in Section~\ref{sec:ext}.

If the segments in $A$ and $B$ and the triangles in $C$ were all full lines
(disjointness then of course cannot be assumed in general, although it might occur when
the lines in each set are parallel, as in the dual version of the 
{\sc GeomBase} problem~\cite{GO-95}), then, in general, determining
the existence of a concurrent triple of lines in $A\times B\times C$, that is,
finding an intersection point of a pair of segments (lines) in $A\times B$
inside a degenerate triangle (line) of $C$,
is the \emph{concurrency testing} problem. This is the dual version of 
the classical \textsc{3Sum}-hard \emph{collinearity testing} problem,
in which we are given three sets of points in the plane, and wish to determine
whether their Cartesian product contains a collinear triple.
This problem has recently been studied 
in the algebraic decision-tree model by Aronov \etal~\cite{AES},
in a restricted version where two of the sets
are assumed to lie on two constant-degree algebraic curves.

The problems studied here can be regarded as other dual versions of
collinearity testing, where restrictions of a different kind are imposed.
As noted by Chan~\cite{Chan}, the additional disjointness properties
that are assumed here make the problem simpler than collinearity testing
(albeit by no means simple), and its solution appears to
have no bearing on the unconstrained collinearity problem itself.
In Section~\ref{app:disc} we comment on the substantial differences 
between this work and the work by Aronov~\etal~\cite{AES}.

Our technique is based on hierarchical cuttings of the plane,
as well as on tools and properties of segment-intersection range searching.
We also use
the so-called Fredman's trick in algebraic-geometric settings, 
in which the problem is mapped into a primal-dual range searching mechanism involving
points and surfaces in $\reals^6$. This reduction exploits the very recent 
multi-level polynomial partitioning technique of Agarwal~\etal~\cite{AAEZ} 
(see also a similar technique of Matou\v{s}ek and Pat\'akov\'a~\cite{MP}). 
Our range searching mechanism of points and algebraic surfaces in higher 
dimensions is a by-product of our analysis, which appears to be 
broadly applicable to other range-searching applications, and we thus 
regard it as a technique of independent interest; 
see, for example, Proposition~\ref{prop:6767} and its proof.

\paragraph{Point location in arrangements.}
An additional key ingredient of our approach involves point location in an arrangement
of lines in the plane (or an arrangement of curves, or in higher dimensions). 
This is of course a well studied problem with several
optimal solutions~\cite{ST}, but we adapt and use techniques that are
handicapped by the requirement that each operation that examines the
real parameters specifying the lines involves at most \emph{three}
input lines. In contrast, the persistent data structure of~\cite{ST}
needs to sort the vertices of the arrangement from left to right, and
each comparison in this sorting is of the $x$-coordinates of a pair of
vertices, which are in general determined by the parameters of \emph{four}
input lines. The persistent data structure method has been used in \cite{AES,Chan} 
for the study of other \textsc{3Sum}-hard geometric problems.
Here we replace this approach
with one that uses solely the relative positions of triples of lines, 
the so-called \emph{order type} of the arrangement. In this approach
each comparison involves only \emph{three} input lines, which, as we show,
eventually leads to improved performance of the algorithm.

In standard settings, separating the order-type computation from the rest
of the processing makes no sense. This is because obtaining the full order-type information
for $N$ lines also takes $\Theta(N^2)$ time. 
This makes the approach based on the order type noncompetitive, as one can 
just do point location in the line arrangement, in the uniform model, with $O(N^2)$ 
preprocessing. Nevertheless, in the applications
considered in this paper (see Sections~\ref{sec:main} and~\ref{sec:ext}), the input lines have
a special representation, which allows us to avoid an explicit construction
of their order type and obtain this information implicitly
in subquadratic time in the decision-tree model.
The rest of the preprocessing, which takes quadratic time and storage in the uniform model, costs nothing in the decision-tree model.
    
The problem of determining whether and how the order type of an arrangement is
sufficient to construct an efficient point-location data structure has,
to the best of our knowledge, never been addressed explicitly.
As we believe that this kind of ``handicapped'' point location will be useful
for other applications (some of which are mentioned in Section~\ref{sec:ext}), 
we present it in some detail in Section~\ref{sec:belg}.
We also present extensions of this
technique to arrangements of constant-degree algebraic curves in $\reals^2$,
and to arrangements of planes or hyperplanes\footnote{%
  For compactness of presentation, we do not single out the case of lines
  in the plane, and obtain it as a special case of hyperplanes in $d=2$ dimensions.}
in higher dimensions, which
will be used in the applications in Section~\ref{sec:ext}.

The algorithm for solving the within-triangle intersection-counting problem
in the algebraic decision-tree model, and, consequently, also of the other
three problems listed at the beginning of this section, is then presented in Section~\ref{sec:main}.
Additional applications of our technique are presented in Section~\ref{sec:ext};
they include: (i)~counting intersections between two sets of pairwise disjoint 
circular arcs inside disks, and
(ii)~minimum distance problems between lines and two sets of points in the plane.

%---------------------------------------------------------------------------------
\section{Order type-based point location in arrangements} \label{sec:belg}
%---------------------------------------------------------------------------------

\paragraph{Order types.}

An arrangement of non-vertical lines in the plane 
(and, later, curves in the plane, or hyperplanes in higher dimension) 
can be described in the following combinatorial fashion. 
We use the notion of an \emph{order type}, defined for a set $L$ of lines
as follows: Given any ordered triple of lines $(\ell_1,\ell_2,\ell_3)$ from $L$, where both 
$\ell_2$ and $\ell_3$ intersect $\ell_1$, we record the left-to-right order of the 
intersections $\ell_1\cap\ell_2$ and $\ell_1\cap\ell_3$ along $\ell_1$; they might coincide.  
The totality of this information gives, for each line in $L$, the left-to-right order of its 
intersections with every other line it meets. For completeness, we assume the existence 
of an ``infinitely steep'' line $\ell_\infty$, placed sufficiently far to the left,
the order of whose intersections with the ``normal'' lines encodes the order of their slopes. 
This information is dual to the perhaps more familiar notion of an order type for 
a set of points in the plane (see, e.g., \cite{GoPo}). 
A higher-dimensional analog of this information involves recording the order in which 
a line that is the intersection of $d-1$~hyperplanes in~$\reals^d$ meets the remaining 
hyperplanes that meet but do not contain it. We also assume a suitable analog of 
the ``infinitely steep line,'' recursively defined over the dimension.

Back in the plane, the permutations along each line of the intersection points with the 
other lines are called \emph{local sequences}~\cite{GoPo84}.
This view allows us to extend the definition of the order type to $x$-monotone curves,
where each pair of curves is assumed to intersect in at most $s$ points, for some constant~$s$.
In this case the order type gives, for each curve~$\gamma$ 
in the collection, the labeled left-to-right sequence of intersection points with the other curves,
where each intersection point is labeled by the triple $(i,j,k)$, where $i$ and
$j$ are the indices of the two curves that form the intersection, and $k$ indicates
that it is the $k$th leftmost intersection point of the two curves. The order type
also includes the vertical order of the curves at~$x=-\infty$. See Section~\ref{sec:curves}
for further details. 

The significance of the order type is that (a) it only records information for $(d+1)$-tuples of objects,
and (b) it contains enough information that lets us construct the arrangement and preprocess it for fast
point location, without having to access further the real parameters that define the objects.

The problem we tackle now is the following: Given the order type of an arrangement, preprocess this information
into a point location data structure. The preprocessing stage is not allowed
to access the actual geometric description of the objects, such as the coefficients 
of the equations defining the lines, curves or hyperplanes, but
can only exploit the discrete data given by the order type. A query,
in contrast, is allowed to examine the coefficients of the few objects that it encounters.

We present two solutions for this problem. First, we show that, for $d$-dimensional 
hyperplane arrangements, for any $d\ge 2$,
the sampling method of Meiser~\cite{M93} (see also \cite{EHKS20}) can be implemented using 
only order-type information (we get the case of lines in the plane as a special case).
Second, we show that for arrangements of $x$-monotone curves in the plane, a simple variant 
of the separating-chain method for point location~\cite{EGS86,LP} can be implemented such that 
only order-type information is used during the preprocessing. % Since the main part of this submission

%----------------------------------------------------------------------
\subsection{Sampling-based approach for hyperplane arrangements} \label{sec:hyper}
%----------------------------------------------------------------------

Let $H$ be a set of $N$ non-vertical hyperplanes in~$\mathbb{R}^d$, where~$d\ge 2$ is a 
fixed constant. We want to construct the arrangement $\A(H)$ induced by 
$H$, where we are only given the order type of $H$. Essentially,
we are given, for each intersection line formed by $d-1$ hyperplanes, the order of
its intersections with the other hyperplanes. (Alternatively, we are given, 
for each simplex $\sigma$ formed by $d+1$ of the hyperplanes, 
the left-to-right order of the vertices of $\sigma$.)
We only require $H$ not to contain vertical hyperplanes.
We do permit more than $d$ hyperplanes to share a point, as well as other degeneracies.
This is indeed a natural scenario for our applications including segment intersection
counting and its related problems.

We briefly sketch the randomized method first proposed by Meiser~\cite{M93} 
and analyzed in detail by Ezra~\etal~\cite{EHKS20}, and show that 
the order type information is sufficient to construct the data structure.

Before considering the point-location structure, we note that the 
order type suffices to construct a discrete representation of the
arrangement $\A(H)$, in which each $j$-dimensional cell of $\A(H)$,
for $j=1,\ldots,d$, stores the set of all $(j-1)$-dimensional cells that
form its boundary (and consequently of all cells, of all dimensions, 
on its boundary), with respective back pointers from each cell to 
all higher-dimensional cells that contain it on their boundary.
This can be done, e.g., by the Folkman--Lawrence topological representation theorem for
oriented matroids~\cite{FL78}, which, roughly speaking, implies that, given the 
order type of~$H$, one can construct a combinatorial representation for the arrangement~$\A(H)$,
consisting of all \emph{sign conditions}. That is, each face $f$ of $\A(H)$ (of any dimension) % $\ge 0$
is encoded by a sign vector $\{-1,0,+1\}^{|H|}$ representing the above/below/on relation of $f$ with respect to
each hyperplane in $H$; see~\cite{BMS-01} for an inductive proof for the planar case,
and its generalization to higher dimensions in~\cite{BKMS-05}. Given this property,
a na\"ive actual construction of the combinatorial representation of $\A(H)$ is easy to derive,
and is free of charge in the decision-tree model, once the order type of~$H$ is computed. 

\paragraph{Preprocessing.}
Given the arrangement $\A(H)$ and a fixed $\eps>0$, we first construct a random sample~$S$
of~$O(\frac{d^2}{\eps} \log \frac{d}{\eps})$ hyperplanes of~$H$; the size of~$S$ does not depend on~$n$.
We then compute a \emph{canonical} triangulation of the arrangement $\A(S)$.
For each face of $\A(S)$, of any dimension $\ge 2$, we use a fixed rule to designate
a \emph{reference vertex} $p$ of this face. For example, we can take $p$ to be 
the lexicographically smallest vertex of the face, with each vertex represented by the
lexicographically smallest $d$-tuple of the indices of the hyperplanes that contain it and whose intersection is a single point.\footnote{%
  When $p$ is chosen as the bottommost vertex, the resulting triangulation 
  is referred to as the \emph{bottom-vertex triangulation}, but in general the order type 
  does not provide us with this information.} 
We then triangulate each face~$f$ of~$\A(S)$
by the \emph{fan} obtained by adding vertex $p$ to each simplex in the triangulations of the lower-dimensional
faces composing the boundary of $f$ and not incident to $p$. Next, we construct the \emph{conflict list}
for each simplex~$\Delta$ of the triangulation, of any dimension,
defined as the set of hyperplanes of~$H$ that \emph{cross}~$\Delta$, i.e., intersect, but not fully contain, it. 
The conflict list can indeed be constructed using only the order type:
Deciding whether a hyperplane $h\in H$ belongs to the conflict list of $\Delta$ amounts
to testing whether there exist two vertices of $\Delta$ that lie on different sides of $h$,
and each such test is an orientation test of the corresponding $(d+1)$-tuple of hyperplanes: $h$ and the $d$ hyperplanes forming the vertex.

From standard results on $\eps$-nets~\cite{HW}, a suitable choice of the constant
of proportionality in the bound on the sample size guarantees that, with high probability,
the conflict list size is not larger than $\eps n$, for each simplex $\Delta$.
It remains to recurse, for each simplex $\Delta$ of the triangulation, with the hyperplanes in its conflict list.
We therefore construct a hierarchical data structure that decreases the number of hyperplanes 
by a factor $\eps$ at each level. This continues until the number of hyperplanes falls below a suitable constant,
at which point we simply store the remaining hyperplanes.

\paragraph{Answering queries.}
Each point-location query returns the relatively open simplex, of the suitable dimension, 
in the canonical triangulation of $\A(H)$ that the query point $q$ lies in.\footnote{%
  For our applications, as well as for the techniques in~\cite{CIO-16,EHKS20} on which we rely, 
  the crucial information is whether the query point~$q$ lies on any of the hyperplanes in~$H$, 
  information that is provided by the point location.}
Queries are answered as follows. First, we locate the simplex~$\Delta$ of the canonical triangulation of $\A(S)$  
containing the query point~$q$. 
Since $d$ is assumed to be constant, $S$ is also of constant size, and so
locating~$\Delta$ can be done in $O(1)$ time.
Next, we recurse in the data structure attached to~$\Delta$. When we reach a leaf of the hierarchy it remains to locate~$q$,
in the sense assumed above,
in the arrangement of a constant number of hyperplanes stored at the leaf.
This, combined with the information collected along the search path, identifies the cell that contains $q$.
The overall number of these recursive steps is $O(\log{n})$, and
thus answering a query costs $O(\log n)$ arithmetic operations,
where the hidden constant\footnote{%
  The value of this constant depends on the storage allocated to the structure. 
  For example, spending $n^{2d\log d + O(d)}$ on storage guarantees query cost of $O(d^4\log n)$~\cite{EHKS20}.}
is polynomial in $d$.
The following lemma summarizes the result. 
%---------------------------------------------------------------------------------
\begin{lemma}
  \label{lem:otpl}
  Let $H$ be a set of $n$ hyperplanes in $\reals^d$, where $d\geq 2$ is a constant.
  Using only the order type of~$H$, we can construct 
  a polynomial-size data structure that guarantees $O(\log n)$-time point location queries in the arrangement~$\A(H)$;
  the implied constant depends polynomially on~$d$. 
  The (polynomial) preprocessing time and the storage of the data structure cost nothing in the decision-tree model.
\end{lemma}
%---------------------------------------------------------------------------------

%----------------------------------------------------------------------------------------
\subsection{Level-based approach for the order type of \texorpdfstring{$x$}{x}-monotone curves in the plane} \label{sec:curves}
%----------------------------------------------------------------------------------------
Let $\Gamma = \{\gamma_1,\ldots,\gamma_n\}$ be a collection of $n$ $x$-monotone 
unbounded constant-degree algebraic curves in the plane, and let $\A(\Gamma)$
denote the arrangement induced by~$\Gamma$. Let $s=O(1)$ denote the 
maximum number of intersections between any pair of curves of $\Gamma$. 
We assume that the curves are in general position:
there are no tangencies between the curves, and $\Gamma$
does not contain vertical lines. 
Recall that the order type of~$\Gamma$ gives us the following information:
\begin{itemize}
\item For each curve $\gamma\in\Gamma$, the left-to-right order of the
      intersection points of $\gamma$ with the other curves of $\Gamma$;
\item The vertical order of the curves at $x = -\infty$.
\end{itemize}
As already mentioned, each intersection point $p$ is labeled by a triple $(i,j,k)$ of indices,
where $\gamma_i$ and $\gamma_j$ are the pair of curves that intersect
at $p$, and $1\le k\le s$ is the \emph{index} of $p$, meaning that $p$
is the $k$th leftmost intersection point of the two curves.
\begin{comment}
  \footnote{%
  If a pair of curves intersect in fewer than $s$ points, the index
  goes up only to the actual number of intersections of the pair.
  The notation used here ignores this non-essential issue, and is
  made to simplify the presentation.}
\end{comment}
Note that the order type tells us whether the
$k$th leftmost intersection point of $\gamma_i$ and $\gamma_j$
lies to the left or to the right of, or coincides with
the $k'$th leftmost intersection point of $\gamma_i$ and $\gamma_{j'}$,
for any quintuple of indices $i,j,k,j',k'$. (Observe that the quintuple
is defined by only three curves.)
However, in this section we are not concerned with the actual construction of
the order type---we simply assume it is given to us in advance.
Such a construction, in a special context that arises in our applications,
is considered when we discuss these applications, 
in Sections~\ref{sec:main} and~\ref{sec:ext}.

Recall that in the query phase we \emph{do} have access to an
explicit description of the curves. We assume a model of computation in which
the following operations can be performed in $O(1)$ time by the query algorithm:
\begin{itemize}
\item Given a query point~$q$ and a curve $\gamma_j$, decide
  whether $q$ lies above, on, or below $\gamma_j$.
\item Given a query point $q$ and an intersection point~$v$ that is
  labeled~$(i,j,k)$, decide if the $x$-coordinate of $q$ is
  smaller than, equal to, or larger than, the $x$-coordinate of $v$.
\end{itemize}
Executing these basic operations is rather easy for lines (where we always have $k=1$). 
When $\Gamma$ contains higher-degree curves, however, executing the second operation is 
more involved. Concretely, comparing~$q$ to an intersection point $v$, labeled as $(i,j,k)$, 
amounts to testing whether a certain quantified
Boolean predicate~$P$ is satisfied. This predicate~$P$ depends on the
real parameters specifying $\gamma_i$ and $\gamma_j$ and on the
coordinates of $q$. It involves $O(k)$ quantified variables that
represent the $k$ intersection points of $\gamma_i$ and $\gamma_j$
to the left of, and including, $v$, and consists of polynomial
equalities and inequalities, whose number depends on $k$, of constant degree 
(which depends on the degree of the curves of $\Gamma$). Still, 
since $k$ and the degrees of the curves are constant, the predicate~$P$ 
has constant complexity. Hence, its validity can be tested in $O(1)$ time, 
for example using the algebraic algorithmic machinery developed in \cite{BPR}.

\paragraph*{Preprocessing.}
Our point-location data structure is based on the separating-chain method 
for planar maps, due to Lee and Preparata~\cite{LP}, which was later refined 
by Edelsbrunner~\etal~\cite{EGS86}. For the case of $x$-monotone curves, 
the separating-chain method is especially easy to implement,
since we can simply use the levels in the arrangement as separating chains.
This allows to carry out the preprocessing using only order-type information,
as explained next.

Observe that a doubly-connected edge list (DCEL) representation~\cite{dBKOS}
of the arrangement~$\A(\Gamma)$ can be constructed in $O(n^2)$ time using only the
given order-type information, without further accessing the real parametric 
representation of the curves. Specifically, the order type gives us the
local sequences of intersection points along each curve, and, assuming for 
the moment general position, we can identify, for each intersection point, 
its four incident edges. Using this data we can trace the boundary of each 
2-face of the arrangement, and consequently obtain the DCEL structure 
(observing that each face is $x$-monotone in this setup).
Recall that we have assumed general position of the curves. Nevertheless, 
if more than two arcs meet at a vertex $v$, we also need to know
the circular order of the incident curves around $v$, which we can 
deduce from the order of the curves at $x = -\infty$ and from the
indices of $v$ along each curve, as we have assumed that all the crossings are proper.
As this is not the scenario that we assume here, we omit further details
of this extension.

Let $\Lambda_j$, for $j=0,\ldots,n-1$, denote the $j$th level in the
arrangement $\A(\Gamma)$, that is, $\Lambda_j$ is the closure of the set
of points that lie on the curves of $\Gamma$ and have exactly $j$~curves
passing below them. Note that the levels can easily be extracted 
from the DCEL of~$\A(\Gamma)$, as the rule for constructing a level is
to follow it from left to right, switching at each vertex to the other
curve forming that vertex. (This latter rule has to be modified, in an 
easy manner, when more than two curves are incident to $v$.) 
We store each level~$\Lambda_j$ as a sorted sequence of its vertices, 
in left-to-right order, where each vertex is
represented as a triple $(i,j,k)$, as explained above
(when more curves are incident to the vertex, any one of the
representing triples suffices).

\paragraph*{Answering queries.}
To answer a query, we perform a binary search on the levels $\Lambda_0,\ldots,\Lambda_{n-1}$.
At each step of this primary binary search we need to decide whether the query point~$q$ 
lies above, on, or below a level~$\Lambda_j$. We can do this by a secondary binary search,
this time on the $x$-coordinates of the vertices of~$\Lambda_j$.
This gives us an edge~$e$ of $\Lambda_j$ intersecting the vertical line through~$q$. 
By comparing $q$ to the curve $\gamma_i\in\Gamma$ defining~$e$, we can determine 
the position of $q$ relative to~$\Lambda_j$. If $q$ lies on $\Lambda_j$
then we are done, otherwise we continue the primary binary search. 
When the query algorithm has finished, we have either identified an edge (or vertex)
of $\A(\Gamma)$ containing~$q$, or an edge immediately above (or below)~$q$. 
Since the DCEL gives us, for each edge~$e$ of $\A(\Gamma)$, the two adjacent faces of $\A(\Gamma)$,
we can now answer the query, returning the face, edge or vertex containing $q$. 

The cost of the search
is $O(\log^2 n)$ (where the constant of proportionality is a large, albeit
constant, function of the degree of the curves of $\Gamma$).  

%----------------------------------------------------------------------
\begin{lemma}
Let $\Gamma$ be a set of $n$ unbounded $x$-monotone constant-degree algebraic curves 
in the plane. Using only the order-type information of~$\Gamma$, we can construct
a data structure that uses $O(n^2)$ storage and that allows us to
answer point location queries in the arrangement~$\A(\Gamma)$ in $O(\log^2 n)$ time. 
\end{lemma}
%----------------------------------------------------------------------

\noindent\textbf{Remarks.} \textbf{(1)}
The query time of the procedure described in Section~\ref{sec:hyper} for $d=2$ (which is $O(\log n)$)
is faster than the time of the procedure presented here for curves in the plane (which is $O(\log^2 n)$).
However, the preceding sampling-based method does not extend to non-straight curves, since there is no 
obvious way to extend the notion of a canonical triangulation to the case of curves. The only viable 
way of doing this seems to use the standard vertical-decomposition technique.
Unfortunately (for us), constructing the vertical decomposition requires that we compare
the $x$-coordinates of vertices defined by different, unrelated pairs of curves. Such a 
comparison involves \emph{four} input curves and it cannot be resolved from the order type 
information alone. For lines in the plane, however, the above technique does yield the
improved logarithmic query time.

\noindent\textbf{(2)} 
We have assumed that the given curves are constant-degree algebraic curves, but the
same machinery would have worked if the curves were arbitrary and we have constant-time
black-box routines that perform the two operations, testing whether a point lies above, 
below, or on a curve, and testing whether a point lies to the left or to the right of 
a specific intersection point of two curves.

\noindent\textbf{(3)} 
It is tempting to apply fractional cascading~\cite{CG} to reduce the
query time to $O(\log n)$. This is problematic in our context, however,
because to implement fractional cascading, we must be able to merge 
suitable sorted subsequences of the sequences of vertices of different levels in the arrangement.
Such a merge requires comparing the $x$-coordinates of two vertices
on different levels, which is not possible using order type only.

%---------------------------------------------------------------------------------
\section{The algorithm for within-triangle intersection-counting} 
\label{sec:main}
%---------------------------------------------------------------------------------
Our input consists of two sets $A$, $B$, each of $n$ pairwise disjoint
segments in the plane, and of a set~$C$ of $n$~triangles in the plane.
To simplify the presentation, we assume that the input is in general
position, namely that, among the segments of $A$, $B$, and edges of triangles
of $C$, no two share a supporting line, and no endpoint of one segment lies on
another (with the obvious exception of the vertices of a triangle in $C$).\footnote{%
  For technical reasons, we also allow a triangle in $C$ to degenerate to a segment.}
These are the only general position assumptions that we need. A triple of segments
(one from $A$, one from $B$, and an edge of a triangle from $C$) are allowed to be concurrent.

\paragraph{A high-level roadmap of the algorithm.}
To avoid various technical issues that complicate the description of our algorithm, we focus in
this overview on the simpler segment concurrency problem, where $C$ is a set of
(not necessarily disjoint) segments, and the goal is to determine whether there
is a triple $(a,b,c)\in A\times B\times C$ of concurrent segments.
To make the overview even simpler, assume that $C$ is a set of lines.

We fix a parameter $g \ll n$ and put $r := n/g$. We construct a 
$(1/r)$-cutting $\Xi(A)$ for the segments of~$A$, and another such cutting
$\Xi(B)$ for the segments of $B$. Since the segments of $A$ are 
pairwise disjoint, we can construct $\Xi(A)$ of size
$O(r)$, and similarly for $\Xi(B)$ (see~\cite{dBS}). We overlay the two cuttings and obtain
a planar decomposition $\Xi$. While the complexity of $\Xi$ is $O(r^2)$,
any line of $C$ crosses only $O(r)$ of its cells.

For each cell $\sigma$ of $\Xi$, we preprocess the sets $A_\sigma\subseteq A$ 
and $B_\sigma\subseteq B$ of those segments that cross $\sigma$, each of size 
at most $n/r = g$, into a data structure that supports efficient queries,
each specifying a line $c$ and asking whether $c$ passes through an 
intersection point of a segment of $A_\sigma$ and a segment of $B_\sigma$.
We pass to the dual plane, obtain sets $A^*_\sigma$ and $B^*_\sigma$ of 
at most $g$ points (dual to the lines containing the segments) each. 
(We ignore here `short' segments that have an endpoint 
inside $\sigma$; see below.) The query is a point $c^*$ and the task is 
to determine whether $c^*$ is collinear with a pair of points
$(a^*,b^*)\in A^*_\sigma\times B^*_\sigma$. For $a\in A_\sigma$ and 
$b\in B_\sigma$ we define $\gamma_{a,b}$ to be the line that passes
through $a^*$ and $b^*$, and let $\Gamma_\sigma$ denote the collection
of these lines. The query with $c^*$ then reduces to point location in
the arrangement $\A(\Gamma_\sigma)$, where we only need to know whether 
$c^*$ lies on any of the lines.

We cannot perform this task explicitly in an efficient manner, since 
the complexity of $\A(\Gamma_\sigma)$ is $O(g^2)$ and we have 
$O(r^2) = O(n^2/g^2)$ such arrangements, of overall quadratic size.
We can do it, though, in the algebraic decision-tree model, in
an implicit manner, using the so-called \emph{Fredman's trick}.
Concretely, we apply the order-type based machinery of 
Section~\ref{sec:belg} to construct $\A(\Gamma_\sigma)$ and 
preprocess it for fast point location.
More precisely, we first construct the order type of $\Gamma_\sigma$: this involves,
for each triple of lines $\gamma_{a_1,b_1}$, $\gamma_{a_2,b_2}$, $\gamma_{a_3,b_3}$, 
determining the ordering of their intersection points along each of these lines. 
We express this test as the sign test of some $12$-variate
constant-degree polynomial $G(a_1,a_2,a_3;b_1,b_2,b_3)$. 

We map the triple $(b_1,b_2,b_3)$ to a point in a six-dimensional parametric space,
and $(a_1,a_2,a_3)$~--- to an algebraic surface $\psi_{a_1,a_2,a_3}$
in this space, which is the locus of all triples $(b_1,b_2,b_3)$ 
with $G(a_1,a_2,a_3;b_1,b_2,b_3) = 0$. 
We now need to locate the points $(b_1,b_2,b_3)$ in the arrangement 
of the surfaces $\psi_{a_1,a_2,a_3}$, from which all the sign tests
can be resolved, at no extra cost in the algebraic decision-tree model,
thereby yielding the desired order type. The subsequent construction 
of the arrangement $\A(\Gamma_\sigma)$, and its preprocessing for fast point 
location, using the machinery in Section~\ref{sec:belg}, also cost nothing in our model.

To make this process efficient, we group together all the points
$(b_1,b_2,b_3)$, over all cells $\sigma$, into one global set $P$, and group the surfaces
$\psi_{a_1,a_2,a_3}$ into another global set $\Psi$. We have
$|P|,\;|\Psi| = O(r)\cdot O(g^3) = O(ng^2)$ (since there are only $O(r)$ cells of
$\Xi(A)$ (resp., of $\Xi(B)$) from which the triples $(a_1,a_2,a_3)$ (resp.
$(b_1,b_2,b_3)$) are drawn).

Using the recent machinery of Agarwal et al.~\cite{AAEZ} (one can also consider 
the alternative technique of Matou\v{s}ek and Pat\'akov\'a~\cite{MP}), we can 
perform this batched point location in 6-space in time 
\[
O\left( |P|^{6/7+\eps}|\Psi|^{6/7+\eps} + |P|^{1+\eps} + |\Psi|^{1+\eps} \right)
= O\left( (ng^2)^{12/7+2\eps} \right) ,
\]
for any $\eps>0$. Full details of this step are given in Section~\ref{app:batched}.

Searching with the dual points $c^*$ takes $O\left(\frac{n^2}{g}\log{g}\right)$ time,
because we have $n$ query lines $c$, each line crosses $O(r) = O(n/g)$ cells 
$\sigma$, and each point location with $c^*$ in each of the encountered arrangements
takes $O(\log{g})$ time. Balancing (roughly) this cost with the preprocessing cost, we choose
$g = n^{2/31}$, and obtain the total subquadratic running time $O(n^{2-2/31+\eps}) = O(n^{60/31+\eps})$.

Quite a few issues were glossed over in this overview. Since the segments
of $A$ and of $B$ are bounded, a cell $\sigma$ may contain endpoints of these 
segments, making the passage to the dual plane more involved. The same applies
in the original segment intersection counting problem, where the triangles of $C$
may have vertices or more than one bounding edge that lie in or meet $\sigma$.
We thus need to handle the presence of such `short' segments and/or `short' triangles.
Moreover, we need to count intersection points within each triangle, and the number of
cells of the cuttings $\Xi_A$, $\Xi_B$ that a triangle can fully contain is much larger 
than $O(r)$. All these issues require more involved techniques, which are developed
below. Still, the overall 
runtime of the resulting algorithm remains $O(n^{60/31+\eps})$, for any $\eps>0$.

\paragraph{Hierarchical cuttings.}
This ingredient is needed for counting intersection points in cells 
that are fully contained inside a query triangle.
Fix a parameter $g \ll n$ and put $r := n/g$. We construct a
\emph{hierarchical $(1/r)$-cutting} $\Xi(A)$ for the segments of $A$, 
which is a hierarchy of $(1/r_0)$-cuttings,
where $r_0$ is some sufficiently large constant. The top-level cutting
$\Xi_1(A)$ is constructed for $A$. Since the segments of $A$ are
pairwise disjoint, we can construct $\Xi_1(A)$ so that it consists of
only $O(r_0)$ trapezoids (for concreteness, we write this bound as
$br_0$, for some absolute constant $b$), each of which is crossed
by at most $n/r_0$ segments of $A$, which comprise the so-called
\emph{conflict list} of the cell $\sigma$, denoted as $A_\sigma$.
The construction time of $\Xi_1(A)$, in the real-RAM model, is
$O(n\log r_0) = O(n)$. See \cite[Theorem~1]{dBS} for details.

\begin{figure}[htb]
\centering
\includegraphics[scale=1.2]{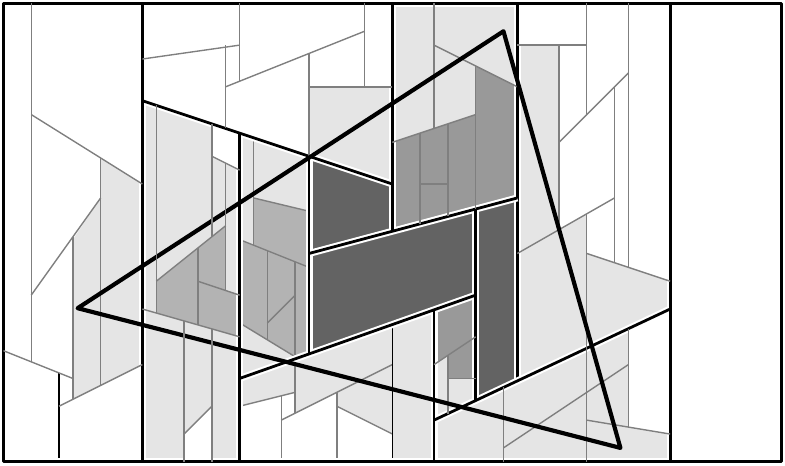}
\caption{\sf Interaction of a hierarchical cutting with a triangle. The dark
grey cells are the ones inside the triangle at the top level
of the hierarchy; the medium grey cells are the ones inside the triangle at
the second level (and whose parent cells are not inside the triangle).
The light grey cells will be refined and handled at lower levels, since
they intersect the triangle boundary.}
\label{fig:hier}
\end{figure}

For each cell $\sigma$ of $\Xi_1(A)$, we clip the segments in its
conflict list $A_\sigma$ to within~$\sigma$ and apply the 
cutting-construction step recursively to this set, clipping also
the cells of the new cutting to within $\sigma$ (and ignoring cells, 
or portions thereof, that lie outside $\sigma$, as they are not met by
any of the clipped segments of $A_\sigma$). We denote the union
of all the resulting $(1/r_0)$-cuttings as $\Xi_2(A)$. We continue
recursively in this manner, until we reach a level $s$ at which
all the cells are crossed by at most~$n/r$ segments. We thus
obtain a hierarchy of cuttings $\Xi_1(A),\Xi_2(A),\ldots,\Xi_s(A)$,
for some index $s = O(\log r)$. We denote the collective hierarchy
as~$\Xi(A)$. Since we stop the recursion as soon as $n/r_0^s \le n/r$,
the overall number of cells of all the levels is $O((br_0)^s) = O(r^{1+\eps})$,
for any prespecified $\eps>0$, for a suitable choice of $r_0=r_0(\eps)$.
Technically, the trapezoids in the cutting are relatively open,
and the cutting also includes one- and zero-dimensional cells; as the latter
are easier to deal with, we will focus below on the two-dimensional cells of the cutting.
At any level $j$ of the hierarchy, the cells of~$\Xi_j(A)$
are pairwise disjoint. As these cells partition the plane, each
intersection point between a segment of $A$ and a segment of $B$
lies in precisely one cell of each level.
See Figure~\ref{fig:hier} for an illustration.

We apply a similar hierarchical construction for $B$, and let
$\Xi(B) = \{\Xi_j(B)\}_{j\le s}$ denote the resulting hierarchical
cutting, which has analogous properties. (We assume for simplicity that
the highest index~$s$ is the same in both hierarchies.)

We now overlay $\Xi(A)$ with $\Xi(B)$, that is, at each level~$j$
of the hierarchy, we overlay the cells of $\Xi_j(A)$ with the
cells of $\Xi_j(B)$. We denote the $j$th level overlay as $\Xi_j$,
and the entire hierarchical overlay structure as $\Xi = \{\Xi_j\}_{j\le s}$.
Since each of $\Xi_j(A)$ and $\Xi_j(B)$ consists of at most $(br_0)^j$ cells,
the number of cells of $\Xi_j$ is at most $O((br_0)^{2j})$. Since we have
$r_0^s \approx r$ (up to a factor of $r_0$), it follows that the overall 
complexity of all the overlays is $O(r^{2+\eps})$, provided that we 
choose $r_0$, as above, to be sufficiently large, as a function of $\eps$.

For simplicity of exposition, we ignore lower-dimensional faces of the cuttings, and
regard each of the overlays $\Xi_j$ as a decomposition of the plane into pairwise
openly disjoint convex polygons, each of complexity linear in $j\le s = O(\log r)$.
Each cell $\sigma$ of the overlay is identified by the pair $(\tau,\tau')$,
where $\tau$ and $\tau'$ are the respective cells of $\Xi_j(A)$ and
$\Xi_j(B)$ whose intersection is $\sigma$;
we simply write $\sigma = (\tau,\tau')$.
Each bottom-level cell $\sigma$ of the
final overlay~$\Xi_s$ is crossed by $\le n/r=g$ segments of~$A$
and by $\le n/r=g$ segments of~$B$.

\paragraph{Classifying the segments and triangles.}
Let $\sigma = (\tau, \tau')$ be a cell of $\Xi_j$, for any level $j$ of
the hierarchy. Call a segment $e$ of $A$
\emph{long} (resp., \emph{short}) \emph{within $\sigma$} if $e$ crosses $\sigma$
and neither of its endpoints lies in $\sigma$ (resp., at least one endpoint
lies in $\sigma$). Let $A_\sigma^l$ (resp., $A_\sigma^s$) denote the set
of long (resp., short) segments of $A$ within~$\sigma$. Apply analogous
definitions and notations to the segments of $B$. Denote by $C_\sigma$
(resp., $C^{(0)}_\sigma$) the set of triangles with at least one edge
that crosses $\sigma$ (resp., that fully contain $\sigma)$.
Call a triangle $\Delta\in C_\sigma$ \emph{long} (resp., \emph{short})
in $\sigma$ if $\sigma$ does not (resp., does) contain a vertex of $\Delta$,
and denote by $C_\sigma^l$ (resp., $C_\sigma^s$) the set of long
(resp., short) triangles in~$C_\sigma$.

For each triangle $\Delta\in C$, each of its edges crosses only $O((br_0)^j)$
cells of $\Xi_j$. Indeed, as such an edge crosses from one cell of $\Xi_j$
to an adjacent cell, it does so by crossing the boundary of either a cell of
$\Xi_j(A)$ or a cell of $\Xi_j(B)$, and the total number of such crossings is
$O((br_0)^j)$. In particular, the edge crosses at most $O(r^{1+\eps})$ cells
of the final overlay $\Xi_s$. It follows that
$\sum_{\sigma \in \Xi} |C^l_\sigma| \le \sum_{\sigma \in \Xi} |C_\sigma| = O(nr^{1+\eps})$,
but clearly $\sum_{\sigma \in \Xi} |C^s_\sigma|$ is only $O(n\log{r})$.
In contrast, $\Delta$ can fully contain many more cells of $\Xi_s$, perhaps
almost all of them, but the hierarchical nature of the construction allows
us to deal with a much smaller number of such interior cells, by collecting
them at higher levels of the hierarchy; see below for details.

\paragraph{The algorithm: A quick review.}
The high-level structure of the algorithm is as follows (see also the `roadmap'
overview given earlier). We construct the hierarchies 
$\Xi(A) = \{\Xi_j(A)\}_{j\ge 1}$ and $\Xi(B) = \{\Xi_j(B)\}_{j\ge 1}$.  
For each cell $\tau$ of $\Xi_j(A)$ (resp., $\tau'$ of $\Xi_j(B)$),
we compute its conflict list $A_\tau$ (resp., $B_{\tau'}$), which,
as we recall, is the set of all segments of~$A$ that cross~$\tau$
(resp., segments of $B$ that cross~$\tau'$). We then form the hierarchical
overlay $\Xi = \{\Xi_j\}_{j\ge 1}$, and for each cell $\sigma = (\tau,\tau')$
of any overlay $\Xi_j$, we compute the subset~$A_\sigma$ of the segments of~$A_\tau$ 
that cross $\sigma$, and the subset~$B_\sigma$ of the segments of~$B_{\tau'}$ 
that cross $\sigma$.  We partition $A_\sigma$ into the subsets
$A_\sigma^l$ and $A_\sigma^s$ of long and short segments (within~$\sigma$),
respectively, and apply an analogous partition to $B_\sigma$. The additional
overall cost for constructing these sets, over all hierarchical levels, is
$O(r^{2+\eps}\cdot n/r) = O(nr^{1+\eps}) = O(n^{2+\eps}/g)$.
(The cost at the bottom level dominates the entire cost over all levels.)

We also trace each triangle $c\in C$ through the cells of $\Xi$ that
are crossed by its edges, and form, for each cell $\sigma$ of the overlay, the list
$C_\sigma$ of triangles of $C$ with at least one edge that crosses $\sigma$.
We partition $C_\sigma$ into the subsets $C_\sigma^l$ and $C_\sigma^s$, as defined
earlier. As we show below, we can handle, in a much more efficient way,
the short triangles of $C_\sigma^s$, as well as the triangles of $C_\sigma^l$
all three of whose edges cross $\sigma$, simply because the overall number of
such triangle-cell interactions is small. We therefore focus on the triangles
of $C_\sigma^l$ that have only one or two edges crossing~$\sigma$.
For triangles with two crossing edges we use a standard two-level
data structure (where in each level we consider only one crossing edge).
This lets us assume, without loss of generality, that each triangle in 
$C_\sigma^l$ is a halfplane. Each of these halfplanes can be represented
by its bounding line, that is the line supporting the appropriate
crossing edge of the triangle. We flesh out the details below.

We also assume, for now, that all the segments of $A_\sigma$
and of $B_\sigma$ are long in $\sigma$ (and so we drop the superscript $l$).
This is the hard part of the analysis, requiring the involved machinery
presented below. After handling this case, we will address the much simpler
situations that involve short segments and/or short triangles (or triangles
with three edges crossing $\sigma$). The cost of handling 
short segments or short triangles within cells is smaller,
even in the uniform model, since the overall number of short objects within 
cells is smaller.

\paragraph{Handling the long segments.}
We preprocess each level $j$ of the overlay, to compute, for each of its
cells $\sigma = (\tau,\tau')$, the number of intersection points between the
(long) segments of $A_\sigma$ and those of $B_\sigma$ (which, due to the clipping,
lie in $\sigma$). This is a standard procedure that involves computing
the number of pairs of segments from $A_\sigma \times B_\sigma$ whose intersection
points with the boundary of $\sigma$ interleave (these are precisely the pairs
of intersecting segments), and can be implemented to take
$O((|A_\sigma|+|B_\sigma|)\log (|A_\sigma|+|B_\sigma|))$ time;
see, e.g.,~\cite{Ag-90}. We store the resulting count at $\sigma$.

Consider a cell $\sigma$, a segment $a\in A_\sigma$, a segment
$b\in B_\sigma$, and a triangle $\Delta\in C_\sigma$. By assumption,
$\Delta$ has only one edge $c$ or two edges $c_1$, $c_2$ crossing~$\sigma$.
When $a$ and $b$ intersect inside $\sigma$,
the intersection lies in~$\Delta$ if and only if the
triple $(a,b,c)$, or each of the triples $(a,b,c_1)$, $(a,b,c_2)$, has
a prescribed orientation, reflecting the condition that the point $a\cap b$
lies on the side of $c$ (or the sides of $c_1$, $c_2$) that contain $\Delta$.
This orientation (or orientations) can be positive, negative, or zero,
depending on the relative order of the slopes of $a$, $b$, and $c$
(or of $c_1$ and $c_2$), and on whether $\Delta$ lies to the left
or to the right of $c$ (or of $c_1$, $c_2$).

For each halfplane $c^+$ that represents a triangle $\Delta\in C_\sigma$
(the halfplane contains $\Delta$ and is bounded by the line supporting the
single (relevant) edge $c$ of $\Delta$ that crosses $\sigma$), we want either
(i) to represent the set of pairs $(a,b) \in A_\sigma\times B_\sigma$ that
have a prescribed orientation of the triple $(a,b,c)$, as the disjoint
union of complete bipartite graphs, or (ii) to count the number of such pairs.
The subtask (i) arises in cases where $\Delta$ has two edges crossing $\sigma$
and is needed for the first level of the data structure, which we query with
the first crossing edge of $\Delta$. The subtask (ii) arises in the second
level of the structure, which we query with the second crossing edge of
$\Delta$, and in cases where only one edge of~$\Delta$ crosses~$\sigma$.

We also count the number of intersections within $\sigma$, in
$O\left((|A_\sigma|+|B_\sigma|)\log (|A_\sigma|+|B_\sigma|)\right)$ time.
As a matter of fact, with a simple modification of the procedure, we can,
within the same time bound, represent the set of all pairs of segments
$(a,b)\in A_\sigma\times B_\sigma$ that intersect each other (inside $\sigma$)
as the disjoint union of complete bipartite graphs, so that the overall size
of their vertex sets is
$O\left((|A_\sigma|+|B_\sigma|)\log (|A_\sigma|+|B_\sigma|)\right)$.
This follows from standard planar segment-intersection range searching machinery;
see, e.g.,~\cite{Ag-90}. In what follows we focus on just one such graph, and to simplify
the presentation we denote it as $A_\sigma\times B_\sigma$, with a slight abuse of notation. 

\paragraph{Preparing for Fredman's trick.}
We use the infrastructure developed by Aronov~\etal~\cite{AES}, with
suitable modifications, but adapt it to the order-type context. 
We preprocess $A$ and $B$ into a data structure that we will then
search with the points dual to the lines supporting the edges of the
triangles of $C$. For each $a\in A$, $b\in B$, we define $\gamma_{a,b}$
to be the line that passes through $a^*$ and $b^*$, where $a^*$ (resp.,~$b^*$)
is the point dual to~$a$ (resp.,~$b$). By our general position assumption, 
$a^* \ne b^*$, so $\gamma_{a,b}$ is well defined.
Let $\Gamma_0$ denote the set of these $n^2$ lines. Our goal in task (ii) is to count,
for each cell $\sigma$ of any of the overlays, for each point~$c^*$ dual to
an edge of a triangle $\Delta\in C_\sigma$, the number of lines of $\Gamma_0$
that lie above~$c^*$, the number of lines that are incident to~$c^*$, and
the number of lines that lie below~$c^*$. In task (i), we want to represent
each of these sets of lines as the disjoint union of a small number
of precomputed canonical sets. This calls for preprocessing the
arrangement $\A(\Gamma_0)$ into a suitable
point location data structure, which we will then search with each~$c^*\in C^*$,
and retrieve the desired data from the outcome of each query.

As in, e.g., \cite{AES}, a na\"ive implementation of this approach
will be too expensive. Instead, we return to the hierarchical partitions
$\Xi(A)$, $\Xi(B)$, and $\Xi$, and iterate, for each level $j$ of the hierarchy,
over all cells $\sigma = (\tau,\tau')$ of $\Xi_j$, defining
$\Gamma_\sigma := \{\gamma_{a,b} \mid (a,b) \in A_\sigma \times B_\sigma\}$.
In principle, we want to construct the separate arrangements $\A(\Gamma_\sigma)$,
over the cells~$\sigma$, preprocess each of them into a point location data structure,
and search, for each triangle $\Delta\in C$, in the structures that correspond to
the cells of $\Xi$ that are either crossed by (at most) one or two edges of $\Delta$,
or fully contained in $\Delta$. This is also too expensive if implemented na\"ively,
so we use instead Fredman's trick, combined with the machinery
developed in Section~\ref{sec:belg}.

We first observe that, for each triangle $\Delta\in C$, finding the cells
$\sigma$ (at any level of the hierarchy) that $\Delta$ fully contains is easy and
inexpensive. We go over the hierarchy of the overlays $\Xi_j$. At the root
we find, by brute force, all the (constantly many) cells of~$\Xi_1$ that $\Delta$
fully contains, and add their intersection counts 
to our output counter. We then
recurse, in the same manner, in the at most $br_0$ cells of $\Xi_1$ that $\Delta$ crosses.
Thus the number of cells we visit is at most
$
O(r_0^2) \cdot \left( 1 + br_0 + (br_0)^2 + \cdots + (br_0)^s \right) = O(r^{1+\eps}),
$
so the overall cost of this step\footnote{%
  It is for making this step efficient that we use hierarchical partitions.
  A single-shot partition would have forced the query to visit $O(r^2)$ such cells,
  which would make it too expensive.}
is $O\left(n r^{1+\eps}\right) = O\left(n^{2+\eps}/g\right)$.

We therefore focus, for each triangle $\Delta$ of $C$, only on the cells that it
crosses (at every level of the hierarchy), and restrict the analysis for now
to cells at which $\Delta$ is long, with at most two of its edges
crossing the cell. Repeating most of the analysis just given,
the number of these cells is~$O(r^{1+\eps})$ (with a smaller constant of
proportionality, since we now do not have the factor~$O(r_0^2)$, as above).

\paragraph{Constructing $\A(\Gamma_\sigma)$ in the decision-tree model.}
Consider the step of constructing $\A(\Gamma_\sigma)$ for some fixed cell $\sigma$.
Following the technique in Section~\ref{sec:belg}, we perform this step using
only the order type of $\Gamma_\sigma$, and we begin by considering the task
of obtaining the order-type information.
That is, we want to determine, for each ordered triple
$(\gamma_{a_1,b_1}, \gamma_{a_2,b_2}, \gamma_{a_3,b_3})$ of lines of
$\Gamma_\sigma$, whether the point
$\gamma_{a_1,b_1}\cap \gamma_{a_2,b_2}$ lies to the left or to the right of
the point $\gamma_{a_1,b_1}\cap \gamma_{a_3,b_3}$.
Let $G(a_1,a_2,a_3;b_1,b_2,b_3)$ denote the $12$-variate polynomial
(of constant degree) whose sign determines the outcome of the above comparison.
(The immediate expression for $G$ is a rational function, which we turn
into a polynomial by multiplying it by the square of its denominator,
without affecting its sign; our general position assumption ensures that
none of the denominators vanishes.)

Once the signs of all expressions $G(a_1,a_2,a_3;b_1,b_2,b_3)$ are determined, 
we can apply Lemma~\ref{lem:otpl}. 
The rest of the preprocessing, which constructs a discrete representation of
the arrangement, say, in the DCEL format, and turns this representation
into an efficient point location data structure, can be carried out
at no cost in the algebraic decision-tree model.

We search the structure with each triangle $\Delta\in C_\sigma$. We may
assume that $\Delta$ is long in $\sigma$ and that
only one or two edges of $\Delta$ cross $\sigma$, as the other 
cases are easy to handle. Assuming further that there is only one such edge $c$,
locating the dual point $c^*$ in $\A(\Gamma_\sigma)$ takes $O(\log g)$
time, as shown in Section~\ref{sec:belg} (noting that $\Gamma_\sigma$
consists of only $g^2$ lines). With suitable preprocessing, locating $c^*$
gives us, for free in our model, the three sets of the lines that pass above $c^*$,
are incident to $c^*$, or pass below $c^*$. The case where two edges of $\Delta$
cross $\sigma$ is handled using a two-level version of the structure; see below
for details. The point location cost now goes up to $O(\log^2 g)$.

Consider then the step of computing the order type of the lines of $\Gamma_\sigma$,
that is, of computing the sign of $G(a_1,a_2,a_3;b_1,b_2,b_3)$, for every triple of
segments $a_1,a_2,a_3 \in A_\sigma$ and every triple of segments $b_1,b_2,b_3 \in B_\sigma$.
To this end, we play Fredman's trick. We fix a bottom-level cell $\tau$ of $\Xi(A)$.
For each triple $(a_1,a_2,a_3) \in A_\tau^3$, we define the surface
\[
\psi_{a_1,a_2,a_3} = \{ (b_1,b_2,b_3) \in \reals^6 \mid
G(a_1,a_2,a_3;b_1,b_2,b_3) = 0 \} ,
\]
and denote by $\Psi$ the collection of these surfaces, over all
cells $\tau$. We have $N := |\Psi| = O((n/g)^{1+\eps}\cdot g^3) = O(n^{1+\eps}g^2)$.
Similarly, we let $P$ denote the set of all triples $(b_1,b_2,b_3)$,
for $b_1,b_2,b_3 \in B_{\tau'}^3$, over all cells $\tau'$ of $\Xi(B)$.
We have $M := |P| = O(n^{1+\eps}g^2)$. These bounds pertain to the bottommost
level of the hierarchy; they are smaller at levels of smaller indices.

We apply a batched point-location procedure to the points of $P$ and
the surfaces of $\Psi$. The output of this procedure is a collection of
complete bipartite subgraphs of $P\times\Psi$, so that, for each such
subgraph $P_\alpha\times \Psi_\alpha$, $G(a_1,a_2,a_3;b_1,b_2,b_3)$ has
a fixed sign for all $(b_1,b_2,b_3) \in P_\alpha$ and all 
$(a_1,a_2,a_3) \in \Psi_\alpha$. This tells us the
desired signs of $G(a_1,a_2,a_3;b_1,b_2,b_3)$, for every pair of triples
$(a_1,a_2,a_3) \in A_\tau^3$, $(b_1,b_2,b_3) \in B_{\tau'}^3$,
over all pairs of cells $(\tau,\tau')\in \Xi(A)\times \Xi(B)$, and these
signs give us the orientation (i.e., the order of the intersection points)
of every triple of lines $\gamma_{a,b}$. That is, we obtain the order type
of the lines. As remarked in Section~\ref{sec:belg}, we may assume
that this also includes the sorting of the lines at $x = -\infty$,
but, for the sake of concreteness, we will address this simpler
task in some detail later on.

\noindent\textbf{Remark.}
This shuffling of the pairs $(a_1,b_1)$, $(a_2,b_2)$, $(a_3,b_3)$
into the triples $(a_1,a_2,a_3)$, $(b_1,b_2,b_3)$, and the treatment
of the first triple as defining a surface in $\reals^6$ and
of the second triple as defining a point in $\reals^6$, is the
realization of Fredman's trick in our context.

%------------------------------------------------------------------
\subsection{The batched point-location step} \label{app:batched}
%------------------------------------------------------------------

We now spell out the details of the batched point location procedure.
It involves points and surfaces in a six-dimensional parametric space, 
and proceeds by using the recent multilevel polynomial partitioning technique of
Agarwal~\etal~\cite[Corollary 4.8]{AAEZ}.
Specialized to our context, it asserts the following result.

%%%%%%%%%%%%%%%%%%%%%%
\begin{theorem}[A specialized version of Agarwal~\etal~\protect{\cite[Corollary 4.8]{AAEZ}}] \label{thm:aaez}
  Given a set~$\Psi$ of $N$~constant-degree algebraic surfaces in $\reals^6$,
  a set~$P$ of $M$~points in~$\reals^6$, and a parameter~$\delta$, with $0 < \delta < 1/6$,
  there are finite collections $\Omega_0,\ldots,\Omega_6$ of semi-algebraic
  sets in $\reals^6$ with the following properties.
  \begin{itemize}
\item
For each index $i$, each cell $\omega\in \Omega_i$ is a connected semi-algebraic
set of constant complexity.
\item
For each index $i$ and each $\omega\in \Omega_i$, at most
$\frac{N}{4|\Omega_i|^{1/6-\delta}}$ surfaces from $\Psi$ cross $\omega$
(meaning, as in the planar setup, that they intersect $\omega$ but do not
fully contain it), and at most
$\frac{M}{4|\Omega_i|^{1-\delta}}$ points from $P$ are contained in $\omega$.
\item
The cells partition $\reals^6$, in the sense that
\[
\reals^6 = \bigsqcup_{i=0}^6 \bigsqcup_{\omega\in \Omega_i} \omega ,
\]
where $\sqcup$ denotes disjoint union.
\item
The sizes of the collections $\Omega_0,\dots,\Omega_6$ are bounded by a function of~$\delta$, and not of
$|P|$ and $|\Psi|$.
\end{itemize}
The sets in $\Omega_0,\ldots,\Omega_6$ can be computed in $O(N + M)$ expected time,
where the constant of proportionality depends on $\delta$, by a
randomized algorithm. For each $i$ and for every set $\omega\in \Omega_i$, the
algorithm returns a semi-algebraic representation of $\omega$, a reference
point inside $\omega$, the subset of surfaces of $\Psi$ that cross $\omega$,
the subset of surfaces that fully contain $\omega$ (for lower-dimensional cells $\omega$),
and the subset of points of $P$ that are contained in $\omega$.
\end{theorem}
%%%%%%%%%%%%%%%%%%%%%%

We compute the partition of Theorem~\ref{thm:aaez}, for a suitable choice of $\delta$,
and find, for each $\psi\in\Psi$, the sets $\omega \in \Omega_i$, over all
$i=0,\ldots,6$, that it crosses, and those that it fully contains.
For each $i$ and $\omega\in \Omega_i$, let $P_{i,\omega}$ denote the set of
points of $P$ in $\omega$, and let $\Psi_{i,\omega}$ denote the set of surfaces of
$\Psi$ that cross $\omega$. We form three complete bipartite graphs
$P_{i,\omega}\times \Psi^0_{i,\omega}$,
$P_{i,\omega}\times \Psi^+_{i,\omega}$, and $P_{i,\omega}\times \Psi^-_{i,\omega}$,
where $\Psi^0_{i,\omega}$ is the set of surfaces that fully contain $\omega$ (and
thus also $P_{i,\omega}$), and $\Psi^+_{i,\omega}$ (resp., $\Psi^-_{i,\omega}$) is
the set of surfaces for which $\omega$ lies fully in their positive (resp., negative)
side, that is, the side at which the corresponding values of $G$ are positive
(resp., negative). As the parameters of the partition are all constant,
the overall size of the vertex sets of these graphs is $O(M+N)$.
To simplify the notation, we refer to the combined size of the vertex sets
of a complete bipartite graph as the \emph{size} of the graph.

For each $i$ and $\omega$, we also have a recursive subproblem that involves
$P_{i,\omega}$ and the subset~$\Psi_{i,\omega}$ of the surfaces that
cross $\omega$. Putting $r_i := |\Omega_i|$, for $i=0,\ldots,6$,
we have, for each $i$ and $\omega$,
\[
|P_{i,\omega}| \le \frac{M}{4r_i^{1-\delta}}\quad\text{and}\quad
|\Psi_{i,\omega}| \le \frac{N}{4r_i^{1/6-\delta}} .
\]
To handle each recursive subproblem, we pass to the dual $6$-space, with
the roles of~$a_1,a_2,a_3$ and of~$b_1,b_2,b_3$ swapped
(such a swap is justified by the complete symmetry of the setup between
the parameters $a_1,a_2,a_3$ and $b_1,b_2,b_3$), and apply, using
Theorem~\ref{thm:aaez}, a similar partitioning. (We now denote the resulting
collections as $\Omega^*_i$ and their respective sizes as $r^*_i$.) We obtain
a second collection of complete bipartite graphs, still of overall size $O(M+N)$,
now with a somewhat larger constant of proportionality, and a new set of
recursive subproblems. Each of these subproblems can be labeled by the pairs
$(k,\omega)$ and $(\ell,\omega^*)$, where $k$ (resp., $\ell$) is the index
$i$ of the primal collection $\Omega_i$ containing $\omega$ (resp.,
dual collection $\Omega^*_i$ containing $\omega^*$).

For each quadruple $((k,\omega),\,(\ell,\omega^*))$, the corresponding primal
subproblem involves at most $\frac{M}{4r_k^{1-\delta}}$ points and at most
$\frac{N}{4r_k^{1/6-\delta}}$ surfaces, which switch their roles when we pass
to the dual, so each cell of the resulting dual partition generates a subproblem
that involves at most $\frac{M}{16r_k^{1-\delta}(r^*_\ell)^{1/6-\delta}}$
dual surfaces (or primal points) and at most
$\frac{N}{16r_k^{1/6-\delta}(r^*_\ell)^{1-\delta}}$ dual points
(or primal surfaces).

We keep flipping between the primal and dual setups in this manner,
until one of the parameters (number of points or number of surfaces) becomes
smaller than some constant threshold $n_0$, which is chosen to be sufficiently
larger than all the (constant) parameters $r_k$, $r^*_\ell$. When this happens, we solve the problem
by brute force, where the running time, and the overall size of the resulting
collection of complete bipartite graphs, are both proportional to the value
of the other parameter (number of surfaces or number of points, respectively).

The primal-dual recursion is applied whenever $M \le N^6$ and $N \le M^6$.
If $M > N^6$ we recurse only in the primal, and if $N > M^6$
we recurse only in the dual. We terminate, as before, when we reach
subproblems where one of the parameters $M$, $N$ becomes at most~$n_0$.

The resulting recursion has the following performance bounds.
%%%%%%%%%%%%%%%%%%%%%%%%%%%
\begin{proposition} \label{prop:6767}
Let $T(M,N)$ denote the maximum possible sum of the sizes
of the complete bipartite graphs produced by the recursive process
described above, over all input sets of at most $M$ points and at most
$N$ surfaces. Then we have
\[
T(M,N) = O\left( M^{6/7+\eps}N^{6/7+\eps} + M^{1+\eps} + N^{1+\eps} \right) ,
\]
for any $\eps>0$, where the constant of proportionality depends on $\eps$.
The same asymptotic bound also holds for the cost of constructing these graphs.
\end{proposition} 
%%%%%%%%%%%%%%%%%%%%%%%%%%%

\noindent\textbf{Proof.}
We fix $\eps$, and show, using induction on $M$ and $N$, that
\begin{equation} \label{eq:6767}
T(M,N) \le A\left( M^{6/7+\eps}N^{6/7+\eps} + M^{1+\eps} + N^{1+\eps} \right) ,
\end{equation}
for a suitable constant coefficient $A$ that depends on $\eps$.
We use $\delta := \eps/2$ in Theorem~\ref{thm:aaez}; to simplify
the calculations a bit, we will work with $\delta$ instead of $\eps$,
so we put $\eps = 2\delta$.

The base cases are when either $M \le n_0$ or $N \le n_0$.
If, say, $M \le n_0$, then we clearly have the `brute force'
bound $T(M,N) \le n_0 N$, which satisfies the bound in (\ref{eq:6767})
if $A$ is chosen sufficiently large. A symmetric treatment holds when
$N \le n_0$. Assume then that (\ref{eq:6767}) holds for all $M' \le M$,
$N' \le N$, where at least one of the inequalities is strict,
for some parameters $M$, $N$ (both greater than $n_0$), and consider
an instance with a set $P$ of $M$ points and a set $\Psi$ of $N$ surfaces.

Assume first that $N^{1/6}\le M \le N^6$.
We consider one phase of the primal decomposition
followed by one phase of the dual decomposition. Fix two pairs
$(k,\omega)$ (in the primal) and $(\ell,\omega^*)$ (in the dual),
and follow the notations introduced above. Apply the induction hypothesis
to the dual subproblem at $\omega^*$. As argued above, this subproblem
involves at most $\frac{M}{16r_k^{1-\delta}(r^*_\ell)^{1/6-\delta}}$ dual surfaces
(or primal points) and at most $\frac{N}{16r_k^{1/6-\delta}(r^*_\ell)^{1-\delta}}$
dual points (or primal surfaces). Hence, by the induction hypothesis,
the contribution of this subproblem to $T(M,N)$ is at most
\begin{multline*}
A\left( \left( \frac{M}{16r_k^{1-\delta}(r^*_\ell)^{1/6-\delta}} \right)^{6/7+2\delta}
\left( \frac{N}{16r_k^{1/6-\delta}(r^*_\ell)^{1-\delta}} \right)^{6/7+2\delta} \right. \\
+ \left. \left( \frac{M}{16r_k^{1-\delta}(r^*_\ell)^{1/6-\delta}} \right)^{1+2\delta} +
\left( \frac{N}{16r_k^{1/6-\delta}(r^*_\ell)^{1-\delta}} \right)^{1+2\delta} \right) .
\end{multline*}
We assume that the numbers $r^*_\ell$ are the same at each primal subproblem. 
We make this assumption for simplicity and clarity of presentation, but it can 
be removed with a more careful analysis.
Multiplying by the number $r_k r^*_\ell$ of subproblems with the indices $k$, $\ell$,
and simplifying the expressions, the contribution is at most
\[
A\left( \frac{M^{6/7+2\delta}N^{6/7+2\delta}}
{16^{12/7+4\delta}(r_k r^*_\ell)^{13\delta/21-4\delta^2}} +
\frac{ (r^*_\ell)^{5/6+2\delta/3+2\delta^2} }{16^{1+2\delta} r_k^{\delta-2\delta^2} } M^{1+2\delta} +
\frac{ r_k^{5/6+2\delta/3+2\delta^2} }{16^{1+2\delta} (r^*_\ell)^{\delta-2\delta^2} } N^{1+2\delta} \right) .
\]
Recall however that we are in the range $M \le N^6$ and $N\le M^6$, so we have
\[
M^{1+2\delta} \le \frac{ M^{6/7+2\delta}N^{6/7+2\delta} }{N^{2\delta}} \qquad\text{and}\qquad
N^{1+2\delta} \le \frac{ M^{6/7+2\delta}N^{6/7+2\delta} }{M^{2\delta}} ,
\]
as is easily checked. The contribution is thus at most
\begin{gather*}
A M^{6/7+2\delta}N^{6/7+2\delta}
\left(
\frac{1}{16^{12/7+4\delta}(r_k r^*_\ell)^{13\delta/21-4\delta^2}} +
\frac{ (r^*_\ell)^{5/6+2\delta/3+2\delta^2} }{16^{1+2\delta} r_k^{\delta-2\delta^2} } \cdot \frac{1}{N^{2\delta}} +
\frac{ r_k^{5/6+2\delta/3+2\delta^2} }{16^{1+2\delta} (r^*_\ell)^{\delta-2\delta^2} } \cdot \frac{1}{M^{2\delta}} \right) \\
< \frac{A}{49} M^{6/7+2\delta}N^{6/7+2\delta} ,
\end{gather*}
provided that $n_0$ (and thus $M$ and $N$) are sufficiently large. Finally, multiplying
this bound by the $49$ possible choices of $k$ and $\ell$, the resulting bound is
at most $A M^{6/7+2\delta}N^{6/7+2\delta}$, thereby establishing the induction
step for this range of $M$ and $N$.

Consider next the case where $M > N^6$. In this case we only work in the primal.
After one level of recursion, for a fixed pair $(k,\omega)$, we get $r_k$ subproblems,
each involving at most $M/(4r_k^{1-\delta})$ points and at most
$N/(4r_k^{1/6-\delta})$ surfaces. Applying the induction hypothesis at each of
these subproblems, the contribution of each subproblem to $T(M,N)$ is at most
\[
A\left( \left( \frac{M}{4r_k^{1-\delta}} \right)^{6/7+2\delta}
\left( \frac{N}{4r_k^{1/6-\delta}} \right)^{6/7+2\delta} +
\left( \frac{M}{4r_k^{1-\delta}} \right)^{1+2\delta} +
\left( \frac{N}{4r_k^{1/6-\delta}} \right)^{1+2\delta} \right) .
\]
Multiplying by the number $r_k$ of subproblems, and simplifying the expressions, we get at most
\[
A\left( \frac{M^{6/7+2\delta}N^{6/7+2\delta}} {4^{12/7+4\delta}r_k^{13\delta/21 - 4\delta^2}} +
\frac{M^{1+2\delta}}{4^{1+2\delta}r_k^{\delta-2\delta^2}} +
\frac{r_k^{5/6+2\delta/3+2\delta^2} N^{1+2\delta} } {4^{1+2\delta}} \right) .
\]
Since $N < M^{1/6}$, the third term is dominated by the second term, provided that
$n_0$ is sufficiently large (recall that we have chosen it to be much larger than
the quantities $r_k$). Using this fact and multiplying by the number, $7$, of
values of $k$, we establish the induction step for this range.

The case $N > M^6$ is handled in a fully symmetric manner, except that we only work in the dual.
The details are fully symmetric to those in the preceding case, and are therefore omitted.

The running time of the procedure obeys the same asymptotic upper bound, which is a consequence
of the fact that the multi-level cells in $\Omega_0,\ldots,\Omega_6$ and their conflict lists
can be computed in $O(M+N)$ time. We omit the easy details.

This completes the proof of the lemma.
$\Box$

\noindent\textbf{Remark.}
Proposition~\ref{prop:6767} can be extended to any dimension $d$, with a similar proof,
to obtain a primal-dual range searching algorithm involving $M$ points and $N$ surfaces
in $d$ dimensions, assuming full symmetry between the points and surfaces, as above.
The running time of the algorithm (in the uniform model) is
$O(M^{d/(d+1) +\eps}N^{d/(d+1) + \eps} + M^{1+\eps} + N^{1+\eps})$, for any $\eps>0$.

%---------------------------------------
\subsection{The rest of the algorithm}

Using a similar and simpler technique, we can sort the lines of each of
the arrangements $\A(\Gamma_\sigma)$, over all cells $\sigma$, at $x=-\infty$.
(Note that this corresponds to sorting them
in reverse order of their slopes.) Here each comparison is between
a pair of lines, say $\gamma_{a_1,b_1}$ and $\gamma_{a_2,b_2}$, and its
outcome is the sign of some constant-degree $8$-variate polynomial (more precisely,
a rational function turned into a polynomial) $H(a_1,a_2;b_1,b_2)$.
Fredman's trick for this setup leads to a batched point location procedure that
involves $O((n/g)^{1+\eps}g^2) = O(n^{1+\eps}g)$ points and
$O((n/g)^{1+\eps}g^2) = O(n^{1+\eps}g)$ surfaces in $\reals^4$. This task
can be accomplished by a considerably simpler variant of the same technique
presented above, whose running time bound is subsumed in the above bound.

In summary, the information collected so far allows us to obtain
the combinatorial structure of each of the arrangements $\A(\Gamma_\sigma)$,
over all cells $\sigma$ of $\Xi$, and subsequently construct an order-type--based
point-location data structure for each of them, at no extra cost
in the algebraic decision-tree model. The overall cost of this phase, in this model, is thus
$O\left((n^{1+\eps}g^2)^{12/7+\eps}\right)$, for any $\eps>0$.
By replacing $\eps$ by some small multiple thereof, we can write
this bound as $O\left((ng^2)^{12/7+\eps}\right)$, for any $\eps>0$.

Fredman's trick, as applied here, separates the handling
of the conflict lists $A_\tau$, over the trapezoids $\tau$ of $\Xi(A)$,
and the conflict lists $B_{\tau'}$, over the trapezoids $\tau'$ of $\Xi(B)$.
For a cell $\sigma = (\tau,\tau')$ of $\Xi$, not all the segments in
$A_\tau$ necessarily cross $\sigma$, so we have to retain (for $\sigma$)
only those that do cross it, and apply a similar pruning to $B_{\tau'}$.
The cost of this filtering step is $O(g)$ for each $\sigma$, for an
overall cost of $O((n/g)^2\cdot g) = O(n^2/g)$, which is a cost that
we are happy to incur as it is subsumed by the cost of searching 
with the elements of $C$, discussed next.

Interpreted in the dual, this step filters out all lines $\gamma_{a,b}$
from $\Gamma_\sigma$ that pass through a (dual) point whose (primal)
segment does not cross $\sigma$, but we also need to filter out lines
$\gamma_{a,b}$, where the corresponding (long) segments $a$ and $b$
do not meet inside $\sigma$ (or do not meet at all). Filtering by
inspecting all pairs $(a,b)$ would be too expensive in the uniform model,
but, fortunately, we can implement this step free of charge in the
decision-tree model. Indeed, consider the complete bipartite graphs
in the compact representation of all the long pairs $(a,b)$ that intersect
inside $\sigma$ (as constructed above). Once this complete bipartite
decomposition is available, we simply keep in $\Gamma_\sigma$ only those lines
that correspond to the edges of these graphs, a step that costs nothing in
the decision-tree model, since it does not incur any extra comparisons among
the input segments. Once this filtration is performed, we can construct
the arrangement of the surviving lines, at no extra cost, and use the
modified arrangement for the point location searches with the elements of $C$,
discussed next.

%%%%%%%%%%%%%%%%%%%%%%%%%%%%%%%%%%%%%%%%%%%%%%%%
\paragraph{Searching with the elements of $C$.}
We now need to search the structures computed in the preceding phase
with the dual features of the triangles of $C$.

Each triangle $\Delta\in C$ crosses only $O(r^{1+\eps}) = O(n^{1+\eps}/g)$ cells of $\Xi$.
As already observed, handling the cells that $\Delta$ fully contains
is simpler and cheaper. (As a matter of fact, this part can be performed
in the real RAM model, and so the main effort is in handling the bottom-level
cells that are crossed by $\Delta$, as described next.)
Recall that, for a cell $\sigma$ that $\Delta$ crosses,
we say that $\Delta$ is \emph{long} (resp., \emph{short}) in $\sigma$
if $\sigma$ does not contain (resp., contains) a vertex of $\Delta$.
There are at most three cells $\sigma$, at the bottom level of the
hierarchy, in which $\Delta$ is short, and we simply inspect all the
$g^2$ pairs of segments in $A_\sigma\times B_\sigma$, and include those
pairs that intersect inside $\Delta$ in our output count for $\Delta$,
for a total cost of $O(ng^2)$. It therefore suffices
to focus on cells in which $\Delta$ is long. $\Delta$ is not processed
at intermediate-level cells at which it is short; it is only processed
as a short triangle at the relevant bottom-level cells.

Let $\sigma$ be such a bottom-level cell. Using the non-planarity of $K_{3,3}$,
it is easy (and standard) to show that there is at most one cell
$\sigma$ that is crossed by all three edges of $\Delta$, so we can handle
these cells as the cells where $\Delta$ is short, with comparable efficiency.
It thus suffices to assume that $\sigma$ is crossed by only one or two edges
of $\Delta$. In the former case we may replace $\Delta$, for
searching within $\sigma$, by the halfplane bounded by the single
edge that crosses $\sigma$, and in the latter case we may replace
$\Delta$ by the intersection wedge of the two halfplanes bounded
by the two edges that cross $\sigma$. In the former situation
we replace $\Delta$ by the point $c^*$ dual to the line supporting the
single crossing edge, and search the point location structure
constructed for $\A(\Gamma_\sigma)$ with $c^*$. In the latter
situation we replace $\Delta$ by the pair of points $c_1^*$, $c_2^*$
dual to the lines supporting the two crossing edges. We prepare
a two-level data structure, where each level is based on the above
point location structure for $\A(\Gamma_\sigma)$, except that the
first level collects its output as the disjoint union of canonical
sets, and the second level counts intersections within the query
triangle. We then search the top level with $c_1^*$ and search
the resulting substructures of the second level with $c_2^*$.

There are $O(nr^{1+\eps}) = O(n^{2+\eps}/g)$ triangle-cell
crossings, each requiring $O(\log^2 g)$ search time, using
(one or two levels of) the above point location data structure for
each such arrangement, for a total of
${\displaystyle O\left( \frac{n^{2+\eps}\log^2 g}{g} \right)}$ time, or
${\displaystyle O\left( (ng^2)^{12/7+\eps} + \frac{n^{2+\eps}\log^2 g}{g} \right)}$ in total.
We (nearly) balance this bound by taking $g = n^{2/31}$, so
the cost of this procedure, in the algebraic decision-tree model, is
$O(n^{2-2/31+\eps}) = O(n^{60/31+\eps})$, for any $\eps>0$.

We next have to handle short segments and short triangles within cells of $\Xi$
(including triangles that have three edges crossing the cell). As might be expected,
this part is less expensive than the handling of long segments and triangles, as we now show. 

\paragraph*{Handling short segments.}

There are two main tasks that we have to implement for short segments:
(i) count the number of intersection points that involve a short segment
and another segment, at all cells of the overlays at all levels, and
(ii) preprocess them so that, for each bottom-level cell $\sigma$,
we can count, for each query triangle $\Delta$, the number of intersection
points with a short segment within $\sigma$ that lie inside $\Delta$.
We start with the first task.

A segment of either $A$ or $B$ can be short in at most two cells,
at each level of the hierarchy. For each cell $\sigma$ at any fixed
level $j$, let $n_\sigma$ denote the number of short segments
(from $A_\sigma\cup B_\sigma$) in $\sigma$, so we have
$\sum_{\sigma \in \Xi_j} n_\sigma \le 4n$. For each cell $\sigma$, the
overall number of segments that cross $\sigma$ is at most $2n/r_0^j$.

Thus, at each cell $\sigma$ at level $j$, we count the number of
intersections between the $n_\sigma$ short segments and the at most
$2n/r_0^j$ other (long or short) segments. Following standard techniques,
such as in~\cite{Ag-90}, this takes
$$
O\left( n_\sigma^{2/3} (n/r_0^j)^{2/3+\eps} + n_\sigma^{1+\eps} + (n/r_0^j)^{1+\eps} \right)
$$
time (also in the uniform model). By H\"older's inequality,
the sum of these bounds over the cells $\sigma$ is at most
\begin{gather*}
O\left( \left( \sum_\sigma n_\sigma \right)^{2/3} (br_0)^{2j/3} (n/r_0^j)^{2/3+\eps}
+ n^{1+\eps} + (br_0)^{2j} \cdot (n/r_0^j)^{1+\eps} \right) \\
= O\left( n^{2/3} \cdot \frac{ b^{2j/3}n^{2/3} }{ r_0^{j\eps} }
+ n^{1+\eps} + b^{2j}r_0^{(1-\eps)j} n^{1+\eps} \right) .
\end{gather*}
Recalling that $r_0^j \le r_0^s = O(r^{1+\eps}) = O(n^{1+\eps}/g)$, this can be upper bounded by
$O(n^{4/3+\eps} + n^{1+\eps}r_0^j) = O(n^{2+\eps}/g)$, for a slightly larger $\eps$
(assuming that $g < n^{2/3}$, as indeed will be the case),
a cost that is subsumed by that of other steps of the algorithm.

Consider next the second task, of counting the number of intersection points
inside a query triangle that involve a short segment, at the bottom-level cells.
The overall number of such intersection points is only $O(ng)$, and we compute all of them
by brute force, and distribute them among the cells. For each bottom-level
cell $\sigma$, let $P_\sigma$ denote the set of these points in $\sigma$,
and let $C_\sigma$ denote, as above, the set of triangles that cross $\sigma$,
with only one or two crossing edges. To simplify the presentation, we only
consider triangle-cell crossings for which the triangle has just one crossing
edge, so it behaves as a halfplane in the cell. The case of two crossing
edges is handled, as above, via a two-level data structure.
Put $M_\sigma := |P_\sigma|$ and $N_\sigma := |C_\sigma|$, for each cell
$\sigma$, and observe that (i) $\sum_\sigma M_\sigma = O(ng)$,
(ii) $\sum_\sigma N_\sigma = O(n^{2+\eps}/g)$, and
(iii) $N_\sigma \le n$ for each $\sigma$.

Applying the standard machinery for halfspace range counting (see \cite{Ag,AE}),
we can count, also in the uniform model, the number of points of $P_\sigma$ that
lie inside (the halfplanes representing) each of the triangles in $C_\sigma$, in time
$O\left(M_\sigma^{2/3}N_\sigma^{2/3+\eps} + M_\sigma^{1+\eps} + N_\sigma^{1+\eps} \right)$,
for each cell $\sigma$. Summing this bound over $\sigma$, using H\"older's inequality,
we get a total of
\begin{align*}
\sum_\sigma & O\left(M_\sigma^{2/3}N_\sigma^{2/3+\eps} + M_\sigma^{1+\eps} + N_\sigma^{1+\eps} \right) \\
& = O\left(n^{1/3+\eps}\right) \cdot
\sum_\sigma M_\sigma^{2/3}N_\sigma^{1/3} + O\left( n^{1+\eps}g + n^{2+\eps}/g \right) \\
& = O\left(n^{1/3+\eps} (ng)^{2/3} (n^{2+\eps}/g)^{1/3}
+ n^{1+\eps}g + n^{2+\eps}/g \right) \\
& = O\left(n^{5/3+\eps} g^{1/3} + n^{1+\eps}g + n^{2+\eps}/g \right) .
\end{align*}
This bound is subsumed in the overall bound on the cost of the other steps of the algorithm,
as described in Section~\ref{sec:main} (again, assuming that $g$ is not too large).

\paragraph*{Handling short triangles and triangles with three crossing edges.}

As we have already noted, the overall cost of this part is $O(ng^2)$. Indeed,
each triangle $\Delta$ is short in at most three cells, at each level of the hierarchy.
However, we need to count intersection points inside a short triangle only at the
bottom-level cells where the triangle is short. For each such cell $\sigma$,
we count for each triangle $\Delta$ that is short in $\sigma$, by brute force,
the number of intersection points inside $\Delta\cap\sigma$, and this indeed
has a total cost of $O(ng^2)$, well below our overall bound.
The same argument applies to triangles with three edges crossing a bottom cell.
 
We remark that the analysis of these parts of the algorithm, which deal with 
short segments or triangles, also applies in the uniform model.

\paragraph*{Putting it all together.}
In conclusion, we finally have:
%%%%%%%%%%%%%%%%%%%%%%%%%
\begin{theorem} \label{thm:main}
Let $A$ and $B$ be two sets each consisting of $n$ pairwise disjoint
segments in the plane, and let $C$ be a set of $n$~triangles in the plane.
We can count, for each triangle $\Delta\in C$, the number of intersection
points of segments of $A$ with segments of $B$ that lie inside $\Delta$,
in the algebraic decision-tree model, at the subquadratic cost
$O(n^{60/31+\eps})$, for any $\eps>0$.
\end{theorem}
%%%%%%%%%%%%%%%%%%%%%%%%%

%%%%%%%%%%%%%%%%%%%%%%%%%
\begin{corollary} \label{cor:three}
We can solve, in the algebraic decision-tree model, at the cost
of $O(n^{60/31+\eps})$, for any $\eps>0$,
each of the problems (i) intersection of three polygons,
(ii) coverage by three polygons, and (iii) segment concurrency, as
listed in the introduction.
\end{corollary}
%%%%%%%%%%%%%%%%%%%%%%%%%

%-------------------------------------------------------------------------
\section{Extensions} \label{sec:ext}
%-------------------------------------------------------------------------

In this section we give a few additional applications of the paradigm developed in this paper.

%-------------------------------------------------------------------------
\subsection{Circular arc intersection counting}
%-------------------------------------------------------------------------

We have two sets $A$, $B$, each consisting of $n$ pairwise disjoint circular arcs
in the plane, and a third set $C$, consisting of $n$ circles in the plane.
Our goal is to count, for each circle $c\in C$, the number of intersection
points of an $A$-arc with a $B$-arc that lie in the interior of $c$.
Denote by $\bar{\gamma}$ the circle containing $\gamma$, for each
arc $\gamma\in\A\cup B$. Put $\bar{A} = \{ \bar{\gamma} \mid \gamma\in A\}$
and $\bar{B} = \{ \bar{\gamma} \mid \gamma\in B\}$.

Using the standard lifting transform, each circle $a\in \bar{A}$ is 
lifted to a `red' plane $a^*$ in $\reals^3$, and each circle $b\in B$ 
is lifted to a `blue' plane $b^*$ in $\reals^3$. 
For each $a\in \bar{A}$, $b\in \bar{B}$, the line $\lambda_{a,b} = a^*\cap b^*$
intersects the standard paraboloid $\Pi$ in at most two points, and 
the lifted images of the at most two intersection points of the arcs
supported by $a$ and $b$ form a subset of zero, one, or two of these points.
Let $P$ denote the set of these points on $\Pi$, keeping only the points
that correspond to actual intersection points of an $A$-arc and a $B$-arc.
We have $|P|\le 2n^2$. Given a circle $c\in C$ (call such circles `green'), 
we want to count the number of points of $P$ that lie below or on the plane $c^*$.

We dualize the setup in $\reals^3$, using the standard duality that 
preserves the above/below relationship, and get a set $P^*$ of at most 
$2n^2$ `red-blue' dual planes (all tangent to $\Pi$). 
The goal is now to locate the points dual to the planes of $C^*$ 
in the arrangement $\A(P^*)$ of the planes of $P^*$. More precisely, 
we want to count how many planes pass below (or through) each query point.

We thus face the problem of point location in a three-dimensional
arrangement $\A(P^*)$ of a set $P^*$ of $O(n^2)$ `red-blue' planes, each 
determined by a red arc in $A$ and a blue arc in $B$. Of course, we cannot
afford the construction of the full arrangement, so we play Fredman's
trick, as in Section~\ref{sec:main}. That is, we construct a $(1/r)$-cutting
$\Xi_A$ for the $A$-arcs, and a $(1/r)$-cutting $\Xi_B$ for the 
$B$-arcs, each of complexity $O(r)$ (which follows since the arcs
in each set are pairwise disjoint), construct the overlay $\Xi$
of these cuttings, and process each cell $\sigma$ of $\Xi$ separately.
We actually need to construct hierarchical cuttings, as in Section~\ref{sec:main}, 
and, at each cell $\sigma$ of the hierarchy of $\Xi$, at any level,
we also need to count the overall number of intersections of $A$-arcs 
and $B$-arcs that lie inside $\sigma$ (this information will be needed
when $\sigma$ is fully contained inside a circle of $C$). 
As before, we classify each arc of $A\cup B$ at a cell $\sigma$ as long 
(resp., short) if the cell does not contain (resp., contains) an endpoint of the arc.
It suffices to focus on long arcs, as short arcs can be handled in much the
same way as in Section~\ref{sec:main}.

As demonstrated in Section~\ref{sec:main},
this subtask is very easy for (long) segments, but is more challenging for circular arcs. 
Still, using the algorithm of Agarwal et al.~\cite{APS}, this can be done,
for the long arcs within each cell $\sigma$, in $O(N_\sigma^{3/2+\eps})$ time, for any $\eps>0$,
where $N_\sigma = |A_\sigma| + |B_\sigma|$, and
$A_\sigma$ (resp., $B_\sigma$) is the set of (long) arcs of $A$ (resp., $B$)
that cross $\sigma$. We have $|A_\sigma|$, $|B_\sigma| \le g := n/r$.

At each bottom-level cell $\sigma$ of the hierarchy of $\Xi$, we need to 
construct, and preprocess for fast point location, the arrangement $\A(P^*_\sigma)$, 
where $P^*_\sigma$ is the set of all dual red-blue planes in $\reals^3$ that are
determined by an arc of $A_\sigma$ and an arc of $B_\sigma$.

We now use the machinery developed in Section~\ref{sec:belg}.
Here we need to perform orientation tests for quadruples of planes
in $P^*_\sigma$, and Fredman's trick allows us to represent each such
test as a truth test of some constant-complexity algebraic predicate 
$G$ in $24$ variables, $12$ variables for the parameters of the four 
circles of $A$ participating in the test, and $12$ variables for the 
parameters of the four circles of $B$. 

In more detail, each plane participating in the orientation test is
dual to a point that is the intersection of~$\Pi$ with some line 
$\lambda_{a,b}$. There are at most two such points, but for such
a point to actually materialize, it needs to belong to the two arcs
$a$, $b$. We assume for now that each of these arcs is long in
$\sigma$ (the other cases are much easier to handle).
We first need to distinguish between the two possible points,
which differ by the sign of the square root in the solution of 
the resulting quadratic equation. Once that is done, and the
four intersection points participating in the orientation
test have been identified, the test itself is the sign test 
of a fixed algebraic expression, but in order to be applicable,
we need to assert that each of the four relevant points $p$
(within the $xy$-plane) lie in $\sigma$ (since we assume that 
our arcs are long in $\sigma$, this suffices to ensure that
the two arcs do indeed intersect at $p$). Put together, all 
these constraints define our desired predicate, which clearly
is of constant complexity.

We transform these tests into point-location
tests of $g^4$ points, formed by quadruples of circular arcs of $A$,
in an arrangement of $g^4$ surfaces, formed by quadruples of circular 
arcs of $B$, in $\reals^{12}$, or the other way around (since we will also 
use duality, in which we flip the roles of $A$ and $B$, so that the 
dual points are determined by arcs of $B$ and the
dual surfaces are determined by arcs of $A$). Again, since the
arcs are assumed to be long, specifying the three real parameters
that define the containing circle, and knowing $\sigma$, uniquely
identifies the arc. 
(The intersection of an arc with a cell $\sigma$ does not have to be 
connected. If it is not connected, we treat each of its connected 
components as a separate arc.)

We group together all these points and surfaces, over the $O(r) = O(n/g)$
(bottom-level) cells of $\Xi_A$ (for the points) and of $\Xi_B$ 
(for the surfaces) into single respective collections $Q$, $\Sigma$, consisting 
of $O((n/g)^{1+\eps}g^4) = O(n^{1+\eps}g^3)$ points and surfaces, respectively. 
(As just mentioned, and as we did in Section~\ref{sec:main}, we use duality, 
so we also treat $Q$ as a collection of dual surfaces and treat $\Sigma$ as 
a collection of dual points in $\reals^{12}$.) Adapting the machinery in 
Section~\ref{app:batched} (see the remark at the end of that section), 
we can solve the latter point-location problem in time
\[
O\left((n^{1+\eps}g^3)^{24/13}\right) = O\left(n^{24/13+\eps}g^{72/13}\right) ,
\]
for any $\eps>0$.
From the output of this procedure we can construct, using the random
sampling machinery in Section~\ref{sec:hyper}, and at no extra cost
in the algebraic decision-tree model, a data structure for point location 
in $\A(P^*_\sigma)$, for each bottom-level cell $\sigma$ of $\Xi$, where
the cost of a query is $O(\log g)$. Arguing as in the preceding section, 
each circle of $C$ has to perform this search at only $O(n^{1+\eps}/g)$ cells,
so the total cost of the point-location searches with the circles of $C$
is $O\left(\frac{n^{2+\eps}\log g}{g}\right)$. Balancing (roughly) this cost 
with the preprocessing cost, we choose $g := n^{2/85}$, and the overall 
cost of the procedure is (the subquadratic bound) $O\left( n^{168/85+\eps} \right)$,
for any $\eps>0$. That is, we have
%----------------------------------------
\begin{theorem}
Given sets $A$, $B$, each of $n$ pairwise disjoint circular arcs in the plane, 
and a set $C$ of $n$ circles in the plane,
we can count, for each circle $c\in C$, the number of intersection
points of an $A$-arc with a $B$-arc that lie in the interior of $c$,
in $O\left( n^{2-2/85+\eps} \right) = O\left( n^{168/85+\eps} \right)$ time, 
for any $\eps>0$, in the algebraic decision-tree model.
\end{theorem}
\subsection{Points and lines in the plane: Minimum distance problems}
%-------------------------------------------------------------------------

In the problem studied in this subsection we have two sets $A$, $B$, each 
of $n$ points in the plane, and a third set $C$ of $n$ lines in the plane, 
and the goal is to determine whether there exist a triple 
$(a,b,c)\in A\times B\times C$, such that $c$ contains a point $x$
that satisfies some property involving $\dd(x,a)$ and $\dd(x,b)$. 
For a concrete example, to be expanded below, given a prescribed 
parameter $t>0$, determine whether any line $c\in C$ contains a point
whose sum of distances to its nearest neighbor in $A$ and its nearest 
neighbor in $B$ is at most $t$. Equivalently, determine whether any $c\in C$ 
intersects any ellipse of major axis $t$ whose pair of foci lie in $A\times B$.

\paragraph*{The problem, in detail.}
Let $A$ and $B$ be two sets, each of $n$ points in the plane, and let $C$
be a set of $n$ lines in the plane. Consider predicates of the form
(where $a$ and $b$ are points, $c$ is a line, and $t$ is a real number)
\begin{equation} \label{eq:pred}
\pi(a,b,c;t):\; \exists x\in c \mid F(\dd(x,a),\dd(x,b)) \le t ,
\end{equation}
where $F$ is a constant-degree bivariate piecewise algebraic function that 
is monotone increasing in both variables, and $\dd$ is the Euclidean distance. 
Typical examples are $F(u,v) = u+v$, $F(u,v) = \max \{u,v\}$, or 
$F(u,v) = u^2+v^2$. Our goal is to determine whether there exist a triple
$(a,b,c)\in A\times B\times C$ such that $\pi(a,b,c;t)$ holds.
For example, when $F(u,v) = \max\{u,v\}$, the goal is to determine whether
there exists a line of $C$ that contains a point that lies at distance at 
most $t$ from a point of $A$ and from a point of $B$.
Similarly, when $F(u,v) = u+v$, the goal is to determine whether
there exists a line $c\in C$ that intersects any `bichromatic' ellipse 
of major axis $t$ that is spanned by a focus in $A$ and a focus in $B$.
Problems of this kind are special instances of facility location, 
where we want to determine whether there exists a line of $C$ that contains
a point whose distance from $A$ and distance from $B$ satisfy some property.
Alternatively, we can aim at reporting all lines of $C$ with this property.

A more ambitious goal (but perhaps not that much more) would be to find the 
minimum value of $t$ for which there exist $(a,b,c)\in A\times B\times C$ such 
that $\pi(a,b,c;t)$ holds, or for which every $c$ has a pair $(a,b)$ such
that $\pi(a,b,c;t)$ holds. For example, for $F(u,v) = u+v$, find the smallest 
major axis of a bichromatic ellipse of this kind that is crossed by some line 
of $C$, or find the smallest major axis of a bichromatic ellipse so that every 
line of $C$ crosses such an ellipse. 

We will consider here only the former setup, in which $t$ is prespecified.
It seems likely that the problem of optimizing $t$ could be solved using parametric search.

The problems studied here can be generalized in several ways, for example by
replacing the lines of $C$ by constant-degree algebraic curves, or by replacing
the Euclidean distance by more general distance functions, but we will deal
only with the problem as formulated above.

For each triple $(a,b,c)\in A\times B\times C$, eliminate $x$ from the
expression in (\ref{eq:pred}) that determines $\pi(a,b,c;t)$, to obtain a 
semi-algebraic region $G(a,b;t)$ in the dual plane (in which lines are 
represented as points), of constant complexity, so that $c\in G(a,b;t)$ 
if and only if $\pi(a,b,c;t)$ holds.

\paragraph*{The algorithm.}
We present a solution for the above problem, that runs in (significantly)
subquadratic time in the algebraic decision-tree model.\footnote{%
  We believe that these are \textsc{3Sum}-hard problems, although we have not yet
  established this property.}
We remark that the problem can be solved in quadratic time
in the uniform model, as will follow from the algorithm that we will derive;
see a comment below to this effect.

By the preceding discussion, we face the problem of point location 
(of the points dual to the lines of $C$) in the planar arrangement 
$\A(\G)$ of the set $\G$ of the `red-blue' regions $G(a,b;t)$, each 
being a semi-algebraic set of constant complexity, and determined by
a red point $a\in A$ and a blue point $b\in B$. As with the previous problems, 
we cannot afford the construction of the full arrangement, so we 
play Fredman's trick, similar to, but in a somewhat different 
context than, the technique in Section~\ref{sec:main}.

We take a random sample $R_A$ of
$r$ points from $A$, and a random sample $R_B$ of $r$ points from $B$, 
construct their Voronoi diagrams $\Vor(R_A)$ and $\Vor(R_B)$, and triangulate 
each cell of the diagrams by triangles emanating from the site of the cell. 
We denote the resulting triangulated diagrams as $\Xi_A$ and $\Xi_B$, 
respectively. Each triangulated diagram has complexity $O(r)$. 
Each cell $\tau$ of $\Xi_A$ (resp., of $\Xi_B$) has an associated 
conflict list~$A_\tau$ (resp.,~$B_\tau$), of those points of~$A$ 
(resp., of $B$) that can be closer to a point in $\tau$ than the
site of $\tau$. With high probability, the size of each conflict 
list is at most $O\left( \frac{n}{r}\log r\right)$. We overlay 
$\Xi_A$ and $\Xi_B$, and obtain a subdivision $\Xi$ of the plane, 
with $O(r^2)$ constant-complexity cells. 
(For problems of this kind, there is no need for a hierarchical 
decomposition, like the ones used in Section~\ref{sec:main} and
in some of the preceding subsections.)

Let $\sigma$ be a cell of $\Xi$, formed by the intersection of a cell
$\tau$ of $\Xi_A$ and a cell $\tau'$ of $\Xi_B$, and let $A_\sigma$
(resp.,~$B_\sigma$) be the points of $A_\tau$ (resp., of $B_{\tau'}$)
that can be closer to points in $\sigma$ than the corresponding sites.

A line $c\in C$ crosses only $O(r)$ cells of $\Xi$. Within each such 
cell $\sigma$, each point $x\in c\cap\sigma$ needs to find its
nearest neighbor $a_x$ in $A$, among the points of $A_\sigma$, and its
nearest neighbor $b_x$ in $B$, among the points of $B_\sigma$, and
then test whether there exists $x\in c\cap\sigma$ such that
$F(\dd(x,a_x),\dd(x,b_x)) \le t$. To do so, within each of these 
$O(r)$ cells, we need to locate the dual point $c^*$ of $c$ in the 
arrangement $\A(\G_\sigma)$, where\footnote{%
  Actually, there is no need to filter away points from $A_\tau$,
  $B_{\tau'}$, to get $A_\sigma$, $B_\sigma$. Keeping all the 
  points from each set does not affect the solution.}
$\G_\sigma = \{G(a,b;t) \mid a\in A_\sigma,\;b\in B_\sigma\}$.
More precisely, we need to determine whether $c^*$ lies in any of these regions.

We now use the machinery developed in Section~\ref{sec:belg}.
Here we need to perform orientation tests for triples of boundary curves 
of the sets $G(a,b;t)$, for $(a,b) \in \Gamma := \bigcup_\sigma (A_\sigma \times B_\sigma)$.
The curves bounding the regions $G(a,b;t)$ are not necessarily $x$-monotone and may
be bounded. This requires some modification of the technique of Section~\ref{sec:curves},
which, albeit technically somewhat involved, are nonetheless rather straightforward
conceptually, and we omit their details, in the interest of brevity.

The construction of the order type of the curves bounding the regions $G(a,b;t)$
amounts to performing various tests, each of which involves three pairs in $\Gamma$,
plus some additional parameters that specify which curves we test and what are the two
intersection points that we compare.
We employ Fredman's trick, which transforms each such test, involving three
pairs $(a_1,b_1)$, $(a_2,b_2)$, $(a_3,b_3)$, to testing whether the point
$(b_1,b_2,b_3) \in \reals^6$ belongs to a certain semi-algebraic region 
$Q_{a_1,a_2,a_3}$, which consists of all points $(u_1,u_2,u_3)$ such that 
$(a_1,u_1)$, $(a_2,u_2)$, $(a_3,u_3)$ satisfy the conditions in the test.
Glossing over some technical issues, this amounts to batched point location
of $O(ng^2)$ points in an arrangement of $O(ng^2)$ surfaces in $\reals^6$.
Applying the machinery in Section~\ref{app:batched}, this can be done in time
$O\left(n^{12/7+\eps}g^{24/7+\eps}\right)$, for any $\eps>0$.
This allows us to construct the arrangements $\A(\G_\sigma)$, over the 
cells $\sigma$, preprocess each of these arrangements for fast point 
location, as in the preceding applications, at no extra cost in the 
decision-tree model. 

As before, searching with the lines of $C$ takes $O\left(\frac{n^{2+\eps}}{g}\log^2 g\right)$
time, and balancing the two costs yields the earlier bound $O\left(n^{60/31+\eps}\right)$,
for any $\eps>0$. That is, we have
%------------------------------------
\begin{theorem}
Let $A$ and $B$ be two sets, each of $n$ points in the plane, let $C$
be a set of $n$ lines in the plane, let $F$ be a constant-degree bivariate 
piecewise algebraic function that is monotone increasing in both variables, 
and let $t$ be a real parameter. We can determine, for each line $c\in C$
whether it contains a point $x$ that satisfies $F(\dd(x,a),\dd(x,b)) \le t$,
where $\dd$ is the Euclidean distance. The algorithm works in the algebraic 
decision-tree model, and takes $O\left(n^{60/31+\eps}\right)$ time,
for any $\eps>0$. 
\end{theorem}
%------------------------------------

% \medskip
\noindent\textbf{Remark.}
As promised, we note that the above algorithm can be adapted, in a much simplified
form, to obtain a quadratic algorithm for the problem in the uniform model.
To do so, we construct the full Voronoi diagrams $\Vor(A)$ and $\Vor(B)$, and form
their triangulated overlay $\Xi$. This step takes $O(n^2)$ time. 
For each cell $\sigma$ of $\Xi$, all its points have the same nearest neighbor 
$a_\sigma$ in $A$ and the same nearest neighbor $b_\sigma$ in $B$. 
Then, for each $c\in C$, we
find the $O(n)$ cells of $\Xi$ that $c$ crosses, and, for each such cell $\sigma$,
we need to test whether $c^*$ lies in $G(a_\sigma,b_\sigma;t)$, an operation that 
takes $O(1)$ time. The overall cost of this step is also $O(n^2)$.

\section{Discussion} \label{app:disc}

As promised in the introduction, we make some comments on the differences
between this work and the work of Aronov~\etal~\cite{AES}, which tackle problems
that have some features in common. Both works use Fredman's trick,
implemented by an offline range-search mechanism, in which objects
in one input set form points and objects in another set form surfaces
in some suitable parametric space. However, the analysis in \cite{AES} works
in the dual plane and uses hierarchical polynomial partitioning for points
(dual to the lines in the input).
This mechanism works efficiently only in the special case where one of the input sets
consists of arbitrary points in the plane, and the other two sets are contained
in \emph{one-dimensional} curves. In this work, we apply a decomposition in the primal
plane (the plane of the segments), and use hierarchical cuttings, where the
crucial property in the analysis is that each set $A$, $B$ consists of
pairwise disjoint segments. This results in a special, low-complexity structure,
which our analysis exploits. In addition, we present a new primal-dual range
searching mechanism, exploiting the recent multi-level polynomial partitioning
technique of \cite{AAEZ}. This mechanism is fairly general and we feel that
it could be used in other range searching applications as well, and is
therefore of independent interest.

Another major difference is the use of order types to construct the
various arrangements $\A(\Gamma_\sigma)$. The fact that each comparison
that we make involves only three objects of $A$ and three of $B$, allows
us to transform it into a test that involves a point and a surface in six
dimensions. In contrast, the standard technique, based on persistent
data structures, calls for sorting the vertices of $\A(\Gamma_\sigma)$
in the $x$-direction, and then each test involves a point and a surface
in eight dimensions. This makes the resulting range searching machinery
considerably less efficient. It is an interesting topic for further research
to find additional applications of this paradigm. 
Puzzlingly, the use of order types seems inapplicable to
the most efficient method presented in \cite{AES}.

Finally, it would be interesting to modify our techniques so as to
obtain (slightly) subquadratic algorithms for these problems in the
uniform model; see Chan~\cite{Chan}.

\paragraph{Acknowledgements.}
We thank Zuzana Pat\'akov\'a for helpful discussions on multilevel polynomial partitioning.

\end{document}